\documentclass{article}
\usepackage[usenames,dvipsnames]{color}
\usepackage{graphicx}
\usepackage{amsfonts}
\usepackage{hyperref}
\usepackage[latin1]{inputenc}
\newcommand\tpc[3]{\begin{topic}[{#2}]\hfill\\\label{#1}{#3}\end{topic}}
\newcommand\tpk[2]{\begin{topic}\label{#1}{#2}\end{topic}}
\newtheorem{topic}{\sf\S}
\newcommand\pc[1]{\S\,{\rm\ref{#1}}}
\newcommand\rmk[2]
{\begin{quote}\label{#1}{\it\small #2}\end{quote}}
\newcommand\REF[1]{{\sf(\ref{#1})}}
\newcommand\RMK[1]{\begin{quote}{\it\small #1}\end{quote}}
\newcommand\CITE[1]{{\sf\cite{#1}}}
\newcommand\rj{\mathbf{r}_\ell}
\newcommand\rjj[1]{\mathbf{r}_{#1}}
\newcommand\sech{\rm sech}
\newcommand\ul[1]{\underline{#1}}

\newcommand\avg[1]{{\left\langle{#1}\right\rangle}}
\newcommand\del[3]{\left(\frac{\partial{#1}}{\partial{#2}}\right)_{#3}}

\newcommand\delp[2]{\frac{\partial #1}{\partial #2}}

\newcommand\dpdz[2]{\frac{d #1}{d #2}}

\usepackage[iso-8859-1]{inputenx}
\title{The Thermodynamics of the Perfect Vapor}
\author{Claudio Zamitti Mammana\\
\normalsize Instituto de F\'{i}sica --- Universidade de São Paulo, Brazil\\
mammana@if.usp.br}
\begin{document}
\cleardoublepage
\maketitle
\vfill \eject
\begin{abstract}
In this paper, heat and Carnot's working substance are defined respectively by the equations $E=3pV$ and $E=3/2pV$, relating the energy $E$, the pressure $p$ and the volume $V$ of these two corpuscular systems. The thermodynamic-kinetic theoretical analysis of their equilibrium reveals that the constraints imposed by the second law on the working substance are the consequence of the relativistic and the quantum taut constraints, respectively, $|v|\le c$ and $|\Delta p\Delta q|\ge h$, imposed on the motion of its molecules. It is further shown that the action of these restrictions is described by the ladder operators.

By revealing the laws that rule the entropy formation mechanism, thereby explaining the time asymmetry, the present approach supersedes statistical mechanics. Furthermore, it establishes the epistemological chain connecting the laws of thermodynamics to the axioms of quantum mechanics, leading to yet another unorthodox interpretation of that theory.
\end{abstract}
\vfil
\begin{center}

Indistinguishability --- Liquid State
\end{center}
\vfil

%1
\section*{Presentation}

\begin{flushright}\small\it
Remove matter entirely from a thing and you remove its capacity for physical change.

{\rm Aristotle}  
\end{flushright}
Looking back at Aristotle's four-element theory we can identify in the opposition of fire against the other three (material) elements, air, water and earth, the fundamental processes that cause their physical changes and, in this opposition, the main subject of thermodynamics and chemistry. After Galileo and Newton, the motion of material particles --- formally acknowledged as a descriptor of the physical changes of bodies --- became the subject of mechanics.

The connection between thermodynamics and mechanics has been hitherto approached by kinetic theory and statistical mechanics. In spite of the revelations provided by quantum mechanics to describe the motion of elementary particles, they still leave unexplained the mechanisms of entropy formation that imply in the time asymmetry in the natural processes.

At the beginning of the last century, it was found that the motion of matter in the universe is constrained by two inequalities, namely, the quantum $|\Delta p\Delta q|\ge h$ and the relativistic $|v|\le c$ relations, which introduce two universal constants into physical theory, namely, the Planck constant $h$ and the velocity of light $c$.

In this paper, I intend to show that it is possible to eliminate the deficiencies of kinetic theory and statistical mechanics if we take the road opened by A. Bartoli and L. Boltzmann, further assuming that the above inequalities describe \textit{taut constraints} that arise when matter interacts with radiation. This approach allows explaining the way in which the molecules of a chemical substance rearrange themselves to accommodate to these constraints, thus engendering its entropy.

The present approach is based on subjecting the Bernoulli \textit{Discrete Fluid} to Bartoli's redefinition of heat. Let us start by reviewing their definitions.

\tpc{DFdef}{Bernoulli Discrete Fluid}{The discrete fluid was defined as a mechanical system of particles (a mechanism) by Daniel Bernoulli in 1738, to stand as an abstract model of the air.}

Bernoulli's approach allowed arriving at the \textit{Fundamental Equation of Molecular Mechanics},
\begin{equation}\label{BEL}
{\cal E}=\frac 32pV.
\end{equation}

Taking into account the revelation by Maxwell that electromagnetic radiation is endowed with \textit{momentum}, {A.\ Bartoli} concluded to be theoretically cogent redefine heat in terms of electromagnetic theory:

\tpc{ABdef}{Bartoli's Redefinition of Heat}{Thermal radiation is, at the same time, a Bernoulli \emph{discrete fluid} \underline{and} a \emph{thermodynamic system}.}

Revelations \pc{DFdef} and \pc{ABdef} allow solve the following propositions: 

\tpk{DotThd}
{To endow the discrete fluid model of \emph{radiation} with thermodynamic faculties.}

\tpk{wsThd}
{To endow the discrete fluid model of \emph{Carnot's working substance} with thermodynamic faculties.}

\RMK{Bartoli's redefinition of heat has two consequences: on one hand, it redefines thermodynamics itself, endowing it with the theoretical elements of the dynamics of particles, thus making of it a \textit{Matter-Radiation Interaction Theory}. On the other hand, it endows the kinetic theory with thermodynamic faculties.}

Although not ordinarily recognized as a ``substance'', electromagnetic radiation was redefined by Bartoli as a Bernoulli Discrete Fluid (a corpuscular substance) endowed with thermodynamic faculties. By applying the thermodynamic methods to the equation,
\begin{equation}\label{pRad}
{\cal E}=3pV,
\end{equation}
Boltzmann could obtain a theoretical derivation of Stefan's radiation law~\CITE{Crepeau} thus solving the proposition \pc{DotThd}.

Perhaps due to the unpretentious style of his presentation, Bartoli's contribution to add thermodynamic predictive power to kinetic theory was largely neglected and fell into oblivion~\CITE{Carazza}, leaving proposition \pc{wsThd} unexplored.

The combination of these two propositions into a single theoretical body of knowledge endows kinetic theory with the analytical power of thermodynamics. Such a program can be easily extended from Newtonian particles to atoms and molecules, thus allowing seeking for a solution for the following proposition:

\tpk{prpA}{To formulate a \emph{Kinetic Theory} capable of explaining, in terms of the theories of mechanics, the variations caused by heat on the thermodynamic properties of its corpuscular model.}

\section*{The Unreality of Classical Mechanics}\label{ArnoldCM}
The state of a real fluid culminates {in the steady state of \textit{equilibrium}}\footnote{As usual in kinetic theory and in statistical mechanics, we adopt the simplifying hypothesis, according to which the discrete fluid has a single equilibrium state. This hypothesis should be reviewed in the case of substances exhibiting a variety of metastable states.}, allowing acknowledge the following empirical evidence:

\tpk{Rot:equipare}
{The final state of the system of particles that compose the discrete fluid is \textit{independent} of its initial state.}

The motion of particles described by classical mechanics is known to be ruled by the following principle:

\tpc{frArnold}
{Principe de D\'eterminisme de Newton}
{L'\'etat inicial d'un syst\`eme m\'echanique (l'ensemble des positions et des vitesses de ses points \`a une date quelconque) d\'efinit de {façon exclusive} le futur de son mouvement.~\CITE{Arnold}.}

The striking contradiction between the empirical revelation \pc{Rot:equipare} and the theoretical principle \pc{frArnold} allows conclude that the laws of classical mechanics cannot explain the motion of gas particles towards equilibrium.

G.\ N.\ Lewis showed that it is not necessary to have recourse to Maxwell's electromagnetic theory to conclude\footnote{\label{ftLewis}See Section \textsl{Radiation and Stefan's Law} in Chapter 10, p. 111~\CITE{Lewis}.} that \textit{heat} exerts pressure on the internal surface of a hollow in contact with a thermal reservoir, thus turning the Bartoli-Boltzmann research program independent of electromagnetic theory. 

To endow a kinetic theoretical model with thermodynamic faculties, the discrete fluid, expressed by the equations \REF{BEL} and \REF{pRad}, will be subjected in Section~\ref{TheBG} to the constraints imposed by the second law of thermodynamics.

\subsection*{The Road not Taken}\label{Road}
The theory prescribed in proposition \pc{prpA} is capable of describing the thermodynamic equilibrium between the phases of a chemical substance in its vapor state.

By traversing the road opened by the Bartoli-Boltzmann program in the seek for the properties of a chemical substance, we will meet the facts and arguments that support the thesis that special relativity has \textit{epistemological precedence} over the kinetic theory of the vapor state.

As much as the thermodynamic analysis of the black body radiation led to the notion of the Planck resonator, the corresponding analysis of the Bernoulli discrete fluid leads to the definition of an abstract representation of Carnot's \textit{working substance}, the \textit{Perfect Vapor}, a two-phase thermodynamic system.

In the course of this presentation, three very old problems are resumed: to explain, in mechanical terms:
\begin{enumerate}\small
\item the temperature, 
\item the asymmetry of time, and 
\item the probabilistic nature of the thermodynamic equilibrium.
\end{enumerate}

\paragraph{\it Temperature.}To deal with phenomena that depend on \textit{temperature}, an exotic theory, \textit{statistical mechanics}, originally formulated by Maxwell and Boltzmann, was refurbished to conform to the quantum restrictions. The most important amendment to that approach was to impose, through the indistinguishability and Pauli's exclusion principles, the symmetry and anti-symmetry of wave functions of many-body systems on the specification of the volume of the phase space. Although successful for the simple cases of the perfect gases of Bose and Fermi particles, this approach does not solve the problem \pc{prpA}.

\paragraph{\it Time.} While \textit{time} is acknowledged as a primitive \textit{independent variable} of classical and quantum mechanics, it is not even considered among the variables that characterize thermodynamics. It is therefore quite remarkable that, among all the theories of physics, thermodynamics is still the only one capable of predicting the time asymmetry observed in nature.

It will be shown that the energy exchanged between the \textit{thermal reservoir} and the \textit{Carnot working substance} occurs during the absorption and emission elementary processes, revealed to be both \textit{stochastically independent}\footnote{The \textit{waiting time between} two consecutive such processes is a continuous random variable.} and \textit{relativistic}. This shows that the role of time in the physics of matter is not primary but determined by a subtle combination of special relativity and quantum mechanics. Hence, time must be acknowledged as a \textit{dependent variable}.

\paragraph{\it Events and their Probabilities.}The \textit{probability} of occurrence of certain predefined \textit{events} became, since Maxwell and Boltzmann, recognized as a necessary element for the establishment of a correspondence between thermodynamics and the mechanical theories.

As stated above, a mechanical \textit{corpuscular system} must also be characterized as a \textit{thermodynamic system}. This double character raises the question: \textit{What dynamic variables are needed to properly characterize a system to meet both these requirements}?

The role of probability in physical theory was reviewed by Max Born in his interpretation of the wave function. Its formal introduction into the equations of quantum theory, however, occurs only after the second quantization, in association with the action of the \textit{ladder operators} on the wave function, expressed in terms of \textit{conditional probabilities}, according to the following statement:

\tpk{TheProb}{The probability of occurrence of a quantum state transition (here associated with the absorption and emission of one {quantum}) by the microscopic state $\ell$ at the instant~$t$, depends on the \emph{random number} $\rj(t)$ of particles in that state at the instant~$t$.}

As I have shown in a previous paper~\CITE{CZM}, the description of the probability distribution function Prob$\left(\rj(t)=k\right)$ for the gases of Bose and Fermi particles, compatible with Born's redefinition of probability in physics, is the Markovian birth and death process, whose \textit{laws of change} derive from the \textit{conditional probabilities} of state transition characteristic of their corresponding ladder operators.

\textit{Stochastic} then supersedes \textit{statistical} mechanics: the physical laws are revealed, not by the state of equilibrium itself, but by those transient processes that lead to equilibrium. This possibility of describing the time evolution of the {\sc pdf} of the random occupation number led to conclude that the indistinguishability of identical particles and the exclusion rule of Pauli are not fundamental principles, but merely conditions of thermodynamic equilibrium. That which we observe in the equilibrium is not an absolute, ontological property of the particles, but the fluctuating statistical indistinguishability.

To give cogency to the explanation of the entropy formation mechanisms, I then had to reinterpret the axioms of quantum theory correspondingly. Although introduced to help explain the particular cases of quantum phenomena involving thermodynamic processes, I hope that these amendments, discussed in the course of this presentation, might throw some light on the interpretation of quantum mechanics.

%2
\section{Thermodynamics of the Discrete Fluid}\label{TheBG}

Since the quantities $p,V$ and $\cal E$ are three independent thermodynamic variables, equations \REF{BEL} and \REF{pRad}, besides describing \textit{discrete fluids}, also define \textit{thermodynamic systems}.

It is known from thermodynamics that these three quantities are related to each other by the differential equation, 
\[ TdS = d{\cal E} + pdV, \]
where $T$ is the temperature and $S$, its entropy.

Differently from the usual practice in kinetic theory, where $\cal E$ is interpreted as the \textit{kinetic energy} of its particles which, according to Bernoulli's explanation, endows the gas with the capacity to exert pressure on the walls of its container, we will interpret ${\cal E}$ in the expressions \REF{BEL} and \REF{pRad} as the \textit{total energy} of the fluid.

By adopting this more encompassing definition of ${\cal E}$, and following \textit{Bartoli-Boltz\-mann program}, it is possible to derive, from \REF{BEL}, the thermodynamic properties of the discrete fluid.

To characterize \textit{both} the Carnot working substance and the heat reservoir as corpuscular and thermodynamic systems, we adopt, respectively, the equations \REF{BEL} \ul{and} \REF{pRad} as their thermodynamic representatives. We can then derive their macroscopic properties by subjecting these equations to the laws of thermodynamics. To ensure generality, we will consider the systems defined by the following common equation:

\begin{equation}\label{eq:kpV}
{\cal E}=\kappa pV,\qquad(\kappa\ \hbox{\rm is a rational number})
\end{equation}
to represent both families of abstract substances.

\subsection{Thermodynamic Properties}\label{Fluide}
To obtain the functional relations involving the thermodynamic variables of these two families of discrete fluids under the constraints imposed by the second law, we subject equation \REF{eq:kpV} to the Maxwell relation\footnote{The conditions imposed on the gas here are essentially the same obtained by Clapeyron in 1834 to derive equation $\frac{dp}{dT}=\frac{\Delta H}{T\Delta V},$ known as the first physicochemical application of the second law of thermodynamics~\CITE{Lewis}.},
\begin{equation}\label{delUVT}
\del{{\cal E}}VT = T\del pTV-p.
\end{equation}

Substituting \REF{eq:kpV} in \REF{delUVT}, we obtain the linear {\sc pde} of the first order,

\begin{equation}\label{eqDifU}
T\del pTV-\frac 1\kappa V\del pVT=\frac 1{\kappa+1}p,
\end{equation}
whose solution is given by any of the following expressions,
\begin{equation}\label{fpVTa}
z = \Phi_p\left(T^\kappa V\right)%, \qquad
=\Phi_V\left(\frac p{T^{\kappa+1}}\right)%, \qquad
= \Phi_T\left[pV ^{(\kappa+1)/\kappa}\right],
\end{equation}
where $z=pV/RT$ and $\Phi_p,\Phi_V,\Phi_T$ are indeterminate functions. Some consequences of this derivation follow:

\tpc{condens}{The Discrete Fluid is a Two-Phase System}{Being $z$ a function of a single variable, according to the phase rule, equations \REF{fpVTa} refer to a two-phase thermodynamic system.}

Let us recall the following theorem\footnote{This theorem, was first stated by Planck in \S\ 52 of~\CITE{PlanckT}.}:

\tpk{I:Lewis}{(A) hollow, at a given temperature, constitutes a simple \textit{thermodynamic system}. If it is brought in contact with a heat reservoir of the same temperature, and its volume is in some way increased or diminished, heat will be taken from or given to the reservoir. The case is \ul{entirel}y \ul{analo}g\ul{ous} to a mixture of liquid and vapor enclosed in a cylinder with a moving piston~\CITE{Lewis}.}

Theorem \pc{I:Lewis} reveals that the discrete fluid exhibits condensation in equilibrium, a phenomenon previewed by Einstein for the gas of Bose particles\footnote{Consider the gaseous phase composed of electrons (fermions) removed by the photoelectric effect from the phase containing the electrons moving inside a metal~\CITE{BridgmanEPM}}.

The following corollary holds for all chemical substances:

\tpk{ThdBP:Potn}{Any form of energy that depends exclusively on the temperature is discarded in the evaluation of the left hand side of equation~\REF{delUVT}, no matter how important it is to the heat capacity of the substance.}

The following corollary derives from \pc{ThdBP:Potn}:

\tpk{RotOsc}{Being functions of temperature only, the molecular energies of rotation and oscillation do not affect the equations of state~\REF{fpVTa}.}

\subsection{The dimensionless requisite}\label{dimless}

Equations \REF{fpVTa} give the arguments of the mathematically arbitrary functions $\Phi_p$, $\Phi_V$ and $\Phi_T$, but not their functional forms. We have some reason to assume that the argument of an arbitrary thermodynamic function cannot be dimensional: in fact, if $\Phi_p$ is, say, a non homogeneous function, $\Phi_p(x)=1/(1-x)$, and its argument $x$, a dimensional quantity such as the length of a straight line segment, then, in its power series expansion, $1+x+x^2+\ldots$, the first term is the pure number 1; the second has the dimension of length; the third the dimension of area; and so on. Whatever unforeseen use one can find for such function, it does not fit the current standards of thermodynamic practice. Hence, to assign physical meaning to the argument of $\Phi_p$, we impose on it the \emph{dimensionless requisite}.

\begin{quote}\small\it %\rmk{Dim-less}
{In the formulation of his hypothesis of the quantum, Planck paid special attention to Dimensional Analysis. With the universal constant $h$, required by imposing the dimensionless requisite on the argument of Boltzmann entropy, independently of its combinatorial character, Planck discovered the possibility of establishing a (non-arbitrary) natural system of units%
\footnote{\S 164 of {\CITE{PlanckT}}},
that acquired importance in cosmological theory, and is being increasingly adopted by many theoretical physicists~\CITE{Kragh}.}
\end{quote}

\begin{quote}\small\it
The additivity of entropy also requires adimensionalty of the argument of the Boltzmann entropy.
\end{quote}

\subsection{A family of abstract thermodynamic systems}\label{Family}

In classical thermodynamics the entropy $S$ of a simple pure substance is expressed as a function $S=F({\cal E},V,m)$ of three variables, namely, its energy ${\cal E}$, the volume $V$ it occupies, and the amount of its substance, here represented by its mass~$m$.

The Boltzmann principle of statistical mechanics, differently, states that the entropy of a physical system {\it in equilibrium\/} depends solely on the (thermodynamic) probability $W$, according to the formula $S=k\ln W$. If statistical mechanics fulfills the requirements to replace equilibrium thermodynamics in physical theory, then these two entropies must be identical:
\begin{equation}\label{SM=Sm}
F({\cal E},V,m)=k\ln W.
\end{equation}

Equation \REF{SM=Sm} cannot be solved unless a detailed microscopic description of the substance is given in terms of the {\it equilibrium\/} occupancy number of quantum states, which, in their turn, are expressed in terms of the quantities ${\cal E},V,m$.

Notwithstanding the indeterminacy of the function $F$, Boltzmann's definition of entropy establishes that the quantity $W$ is an integer pure number. We can then conclude that the argument of $F$ is correspondingly a single, dimensionless, quantity. Hence, to assure mathematical consistency with the Boltzmann principle, we impose on the argument of the function $F$ the \emph{dimensionless requisite}.

To treat our problem by the methods of classical thermodynamics, it is advisable to introduce the abstract prototype $\Theta$ of a family of thermodynamic systems, whose entropy is given by the following definition,

\tpc{tVap}{The Family $\Theta$ of Thermodynamic Systems}{The entropy of a member of the family $\Theta$ of thermodynamic systems is given by a function $F(W)$ of the single dimensionless quantity $W$, given by the monomial,
\begin{equation}\label{UVm}
W = {\cal E}^\chi V^\beta m^\gamma\varsigma.
\end{equation}
where $\chi,\beta$ and $\gamma$ are rational numbers, and $\varsigma$ is a dimensional constant, characteristic of the system, that renders~$W$ dimensionless.}

Once $\varsigma$ is fixed, the values of $\chi, \beta$ and $\gamma$ can be obtained by the methods of Dimensional Analysis. According to the chain rule for the derivative of a function, the entropy is related by the fundamental differential coefficients,
\begin{equation}\label{SVm}
\begin{array}{l}
\del S{\cal E}{V,m}=
\frac 1T=\frac{\chi}{{\cal E}} WF^\prime\left(W\right),
\\
\del SV{{\cal E},m}=
\frac pT=\frac\beta V WF^\prime\left(W\right).
\end{array}
\end{equation}
Dividing the latter by the former and adopting the notation $\kappa=\chi/\beta$
we obtain the general energy law \REF{eq:kpV} of the members of the family $\Theta$, that confirms that radiation (a gas of Bose particles) and the gas of fermions are members of the same family~$\Theta$.

\begin{quote}\small\it %\rmk{R:symm}
The establishment of the functional form of the magnitude $W$ in \REF{UVm} cannot be obtained by purely thermodynamic methods. This issue will be treated in Sections~{\textbf{\ref{EntropyChng}}}, and~{\textbf{\ref{theBlackBodyRad}}}.
\end{quote}

\subsection{Ideal substances defined by universal constants}\label{Ctes}

Although the quantity $\varsigma$ can be any physical constant, we will concentrate here on the already known universal constants, and focus on those connected with thermodynamic phenomena, as revealed by the theories of thermal radiation, quantum mechanics, and electromagnetism\footnote{Although the gravitational constant $G$ can be included among the fundamental constants of physics, considerations about gravitation are beyond the scope of this paper.}, namely Planck's constant~$h$, the speed of light~$c$, and the charge of the electron~$e$.

The family $\Theta$ defined by these constants is subdivided into two main groups, according to the dependence of entropy on the mass of the particles, which reflect the different behaviors determined by the equations \REF{BEL} and \REF{pRad}.

\subsection{Mass-independent entropy}\label{WienK}

Kirchhoff's laws of black-body radiation and Wien's \textit{Displacement Law} contributed to add \textit{spectroscopy} to thermodynamics, which imposes the \emph{Frequency Matching Principle}\footnote{To be introduced in Section~\ref{theBlackBodyRad}.} to the analysis of matter-radiation phenomena. Such revelation introduced another difficulty in the determination of the thermodynamic properties of chemical substances.

\begin{quote}\small\it
In a few cases of abstract substances (the gases of Bose and Fermi particles), these properties could be derived from the introduction of strongly simplifying hypotheses about their spectra, to obtain their \emph{Partition Functions} by the methods of \emph{Operational Calculus}.
\end{quote}

\tpc{BG:nm}{Mass Independent Behavior}{It is known that the electric charge $e$ is connected to the constant $hc$ by the \textit{fine structure constant} $\alpha$. For our purposes, we are then allowed to write $u_{hc}=2\pi\alpha u_e$. The quantity $u_{hc}$ characterizes a gas of massless particles (photons), and $u_e$, a gas composed of electrically charged particles whose behavior does not depend on their masses. Both these gases lead to $\kappa=3$.}

The partition functions of these substances, thereby their entropies, are known to be expressed in terms of a power series expansion of the Boltzmann factor, $e^{-\epsilon_\ell/kT}$, where $\epsilon_\ell$ denotes the element $\ell$ of the energy spectrum of the gas, expressed in terms of its radiation spectrum wavelengths,
\begin{equation}\label{lambdaProp}
\lambda_\ell\propto\sqrt[3]{\frac VN}.
\end{equation}
The lengths $\lambda_\ell$ were assumed by Wien in his derivation of the \textit{displacement law} to be integer divisors of the linear dimensions of the geometry of the container.

\subsection{Mass-dependent entropy}\label{RelatVarS}

Chemical reactions involve relativistic changes of the masses of the particles of the reactants and products\footnote{While this fact has been long admitted in the treatises on chemistry, it has been neglected, either due to the undetectable variation of mass, or due to the then-unknown mechanisms of mass-energy relativistic interconversion processes, previewed in Section~\ref{SecSR}.}. Mass relativistic variation occurs in any system whose entropy depends on the masses of its constitutive components.

{Gases defined in terms of $\varsigma=h$ lead to $\kappa=3/2$. It is not possible to render dimensionless the argument $T^{3/2}V$ of the function $\Phi_p$ in \REF{fpVTa} unless we have recourse to Planck's constant. We therefore adopt the \textit{quantum magnitude},
\begin{equation}\label{dgd}
\theta=\frac h{\sqrt{2m\pi kT}}\sqrt[3]{\frac NV},
\end{equation}
where $m$ is the mass of a molecule of the gas, and $N$ is the number of molecules of the substance contained in the volume $V$. The magnitude $\theta$ can be recognized as the cubic root of the \emph{degree of gas degeneration}, which is ubiquitous in statistical mechanics.}
 
It will be shown that the behavior of these gases depends on the relativistic variation of the masses of their particles, so that their thermodynamic functions depend, not on the Boltzmann factor $\exp\left(-\frac{\epsilon}{kT}\right)$, as hitherto assumed by many authors, but on the magnitude~$\theta$.

\begin{quote}\small\it %\rmk{R:dual}
{Substances composed of electrically charged material particles might exhibit a dual behavior, whether characterized by $\varsigma=e$, or by $\varsigma=h$. The latter occurs when the laws that cause the changes in the state of motion of the particles are relativistic.}
\end{quote}

%3
\section{The Perfect Vapor}\label{TheVap}

\begin{flushright}\small\it
Since, according to (Kirchhoff's) law, we are free to choose any system\\whatever, we now select from all possible \emph{\ul{emittin}g and \ul{absorbin}g\\systems} the simplest conceivable one, namely, one consisting of a\\large number $N$ of similar stationary oscillators \textsl{(\ldots)}~\CITE{PlanckT}% p136
\end{flushright}

It is important to emphasize that, in modeling the black-body radiation, the choice of the \textit{system of resonators} as its representative is not imperative, but was a mere \emph{choice} made by Planck to simplify the derivation of the chromatic distribution function.

\RMK{It is remarkable that the description of the radiation as a system of harmonic oscillators makes no reference to any preliminary hypothesis about the \textit{quantum nature} of the processes involved in the formation of the thermodynamic equilibrium of radiation.} 

In a certain sense, it was an unfortunate historical coincidence that the energy distribution of photons is isomorphic to that of a system of \textit{Planck resonators}\footnote{In my first reading of Einstein's 1917 paper, I became confused when I realized that he was deriving, not the distribution of energy by the molecules of a chemical substance, but, instead, that he was treating of the black-body radiation, itself. We will see that it is not the distribution of energy, but the distribution of quanta, that is, mechanical action, which will show that my first reading was also justified.}, for it led several pioneers of statistical mechanics to confuse the distribution of energy by atoms and molecules of gases with that which holds for photons.

This confusion is still found in several situations: in some formulations of the partition functions of real gases; in the incorrect adoption of Gibbs' canonical ensembles approach to systems involving relativistic phenomena, in particular, chemical reactions; and specially in the derivation of the ladder operators, for it gives to the quantum oscillator an interpretation in quantum physics that concealed the relativistic nature of the physical processes triggered by these operators.

\subsection{The Perfect Vapor}\label{TheVapor}
Recall that Lewis's radiation pressure theorem$^{\rm\ref{ftLewis}}$ justifies adopting the thermodynamic system defined by the equation~\REF{BEL} as an abstract model of the vapor state of chemical substances. The equation of state of this abstract substance (the \textit{perfect vapor}) can be written in terms of the undetermined function, $\phi_p$ of the only argument~$\theta$,
\begin{equation}\label{eos:Phip}
z = \phi_p (\theta).
\end{equation}

\subsection*{Asymptotic Behavior of the Perfect Vapor}%\label{Asympt}
Note that the limit $ \theta\rightarrow 0$ is usually invoked in kinetic theory as the condition under which molecules of a gas can be treated as classical particles, with fairly well-defined position and momentum~\CITE{Huang}. To give more realism to the equation\REF{eos:Phip}, we subject it to the following asymptotic condition:

\tpc{AsymptBehav}{Asymptotic Behavior of a Chemical Substance}{The behavior of all known chemical substances in the thermodynamic equilibrium state, for $\theta\rightarrow 0$, is given by the ideal gas law, $z=1$.}

From \pc{AsymptBehav}, we can express $\phi_p(\theta)$ in terms of the power series expansion,

\begin{equation} \label{eq:State}
z=1-\theta\left(a_1+a_2\theta+a_3\theta^2+\ldots\right)
=1 - f_\sigma(\theta),
\end{equation}
where the $a_i$'s are constants and $f_\sigma(\theta)$ is a function that characterizes the indeterminate substance $\sigma$ of the single argument $ \theta$. Expressed in terms of thermodynamic variables, it acquires the following form,
\begin{equation} \label{pV=RT}
pV+RTf_\sigma(\theta)=RT,\qquad f_\sigma(0)=0,
\end{equation}

The indetermination of the function $f_\sigma(\theta)$ suggests that each substance (or family of substances) $\sigma$ has its own characteristic function.

\subsection*{The Entropy of the Perfect Vapor} % \label{SofBG}
The derivative of the entropy \pc{tVap} for the case of the relativistic variation of the mass of the particles, reveals the existence of a certain type of partial entropy, \textit{hitherto ignored}, which leads to the departure $f_\sigma(\theta)$ from the \textsc{eos} \REF{eq:State}.

Let us denote the partial entropies of the gaseous and liquid phases, respectively, by the symbols $\mathfrak{G}$ and $\mathfrak{L}$. As shown above, the derivation of the thermodynamic properties of the perfect vapor, based on Maxwell's relation \REF{delUVT}, does not preview the expressions of the partial entropy of the liquid phase. To separate these partial entropies we subject the equation of state~\REF{pV=RT}, to another of the Maxwell relations, namely,
\begin{equation}\label{delSVT}
\del SVT = \del pTV,
\end{equation}
which leads to a different \textsc{pde}, whose solution, $S= S_{\mathfrak{G}}+S_{\mathfrak{L}}$,
is given by the following equations,
\begin{eqnarray}\label{Svap}
S_{\mathfrak{G}} &=&-3R\ln\theta,\label{Sc}\\
S_{\mathfrak{L}}&=&-\frac 32Rf_\sigma(\theta)
+ 3R\int\frac{f_\sigma(\theta)}\theta d\theta.\label{Sq}
\end{eqnarray}
The magnitude $S_{\mathfrak{G}}$ can be immediately identified with the Sackur--Tetrode entropy of the \textit{monatomic ideal gas}.

\RMK{In Appendix \ref{Steam}, the \textsc{eos} \REF{eq:State} is confronted against the steam $pVT$ data.}

\subsection*{Radicles and Clusters}
To describe the phases of the perfect vapor, we will use the notions of \textit{radicle} and \textit{cluster}, defined as follows:

\tpc{radicles}{Radicle}
{Let's assume that during the gas-to-liquid conversion processes, or \textsl{vice versa}, the perfect vapor molecules \textit{do not dissociate themselves into smaller molecules}. Instead, during these phase transition processes, they behave as \emph{indivisible units}, as it happens with the atoms in the molecules in chemical reactions. Molecules with this chemical behavior were named \emph{radicles} by Berzelius~\CITE{Singer}.}

\tpc{clusters}{Cluster}
{A cluster $\ell$ is a molecule composed of a random number $\rj(t)$ of radicles. The probability distribution functions which describe the population dynamics of the liquid phase is determined by the conditional probability laws governing the elementary quantum processes of \textit{creation} and \textit{annihilation} of quanta either in radiation (the thermal reservoir) or in the clusters of the liquid phase of the perfect vapor.}

%4
\section{The Matter-Radiation Quantum Exchange}\label{MRProblem}

Maxwell thus described the motion of particles in a gas:

\tpk{MaxDescr}{\it ``During the great part of their course the molecules (of the gas) are not acted on by any sensible force, and therefore move in straight lines with uniform velocity. When two molecules come \emph{within a certain distance of each other, a mutual action takes place between them} {\em (\ldots)}. Each molecule has its course changed, and starts a new path''~{\CITE{Maxwell}}.}

In order to give more realism to Maxwell's description, we must remember that, in the analysis of the thermodynamic properties of the gas, one must consider that the movement of its particles takes place under the action of heat. We then realize that, so formulated, this motion is an instance of the phenomenon treated by Einstein in his \textsl{Electrodynamics of Moving Modies}. In other words:

\tpk{HypoSpecRel}{The matter-radiation interaction is relativistic.}

In this Section, by focusing on the equilibrium of the two phases of the perfect vapor, we will demonstrate conjecture \pc{HypoSpecRel}, recalling that the argument $W$ of the Boltzmann entropy \REF{SM=Sm} has the dimension of action, and taking into account the following hypothesis\footnote{This assumption will be justified and detailed in~\pc{TwoElProcs}.}: 

\tpk{AbsEmiss}{The two elementary energy conversion processes --- absorptions and emissions --- are stochastically independent.}

\subsection{The Principle of Conservation of Action}\label{CvtvAct}
In thermodynamics the vapor state has been approached by the following theorem~\CITE{Lewis}:
	
\tpc{LewisdS}{The Entropy During a Reversible Condensation}
{A system, or any part of a system, undergoes an increase in entropy as it absorbs heat from the medium, resulting in an equal decrease in the entropy of the medium, and that the increase in entropy is equal to the heat so absorbed divided by the absolute temperature,
\begin{equation}\label{1111}
dS =\frac{\delta q_{rev}}{T}.
\end{equation}}

Let us consider a thermodynamic system composed of the perfect vapor in thermal contact with its surroundings (a heat reservoir),
\begin{equation}\label{uhc=theta}
\hbox{\rm Radiation + Perfect Vapor}.
\end{equation}
According to the definition \pc{tVap}, both the radiation and the perfect vapor are members of the family $\Theta$, whose entropies are given, respectively, by the indeterminate functions, $F\left(u_{hc}\right)$, and $f(\theta)$.
	 
Moreover, from the point of view of Bartoli thermodynamics, radiation is a thermodynamic system in its own right, so that the equilibrium between matter and radiation presupposes both the separate equilibrium of radiation and matter, and the equilibrium of a three-phase system, conjointly,
\begin{equation}\label{3phases}
\hbox{Radiation + Gaseous Phase + Liquid Phase}.
\end{equation}

Recall that the equilibria of the radiation (the heat reservoir) and of the perfect vapor, treated separately, are characterized by the stationary statistical fluctuation of the quantities $u_{hc}$ and $\theta$, respectively. Hence, according to theorem~\pc{LewisdS}, the equilibrium of the combined system \REF{3phases} is achieved when, for each degree of freedom, we have, $u_{hc}=\theta$, or
\begin{equation}\label{preut}
\frac{\nu}{kT} = \frac 1{\sqrt{2\pi mkT}}\sqrt[3]{\frac NV}.
\end{equation}

Replacing the proportionality relation \REF{lambdaProp} in \REF{preut} we get the expression,
\begin{equation}\label{uhctheta}
{kT}\propto mc^2.
\end{equation}

The relation \REF{uhctheta} together with the conjecture \pc{HypoSpecRel} justify assuming that the heat-work interconversion processes that characterize thermodynamics are relativistic. We are then allowed to (provisionally) assume the scalar relation as the description of this process:
$\Delta\epsilon\propto\Delta mc^2$.
% ifdefs [202]

\subsection{Relativistic Character of Equilibrium}\label{SecSR}
It is then natural to assume that the energy converted during an elementary thermodynamic transformation occurring in a material substance in the presence of an electromagnetic field (a thermodynamic system in contact with a heat reservoir) can be obtained, not from a classical approach, but by from the following relativistic equation,
\begin{equation}\label{e2}
\epsilon^2=m^2c^4+c^2p^2.
\end{equation}

In addition to the well-known factorization of Dirac's expression in four dimensions of the expression,
\[\sqrt{m^2c^4+c^2p^2}, \]
there is another, more elementary one, which, despite its importance, has been neglected by kinetic theory,
\begin{equation}\label{ee}
\epsilon^2% = m^2c^4+c^2p^2
= \left(mc^2+\imath cp\right)\left(mc^2-\imath cp\right).
\end{equation}
The equation \REF{ee}, interpreted in terms of the Minkowski representation, which gives physical meaning to the imaginary unit $\imath$, acquires physical significance in the following circumstances:

When a material particle of mass $m$ interacts with electromagnetic radiation, its energy undergoes changes caused by the action of the elementary energy conversion processes\footnote{Such inter-conversion processes can be identified with radiation absorption and emission.} occurring within chemical substances. {Due to the assumption~\pc{AbsEmiss}}, the energies $\Delta\epsilon$ or $\Delta\epsilon^*$ can be expressed in terms of \textit{two independent complex numbers}, whose values are given, respectively, by the equations,
\begin{eqnarray}
&\Delta\epsilon   &= \Delta mc^2+\imath c\Delta p,\label{+Deltap}\\
&\Delta\epsilon^* &= \Delta mc^2-\imath c\Delta p.\label{-Deltap}
\end{eqnarray}

This proves \pc{HypoSpecRel}.

\RMK{The conjecture \pc{HypoSpecRel} implies reformulate Maxwell's description \pc{MaxDescr}:}

\tpc{MinkS4}
{Maxwell's description in Minkowski space}
{{\rm(\ldots)} when two molecules come within a certain \emph{interval of each other in the timeless Minkowski space phase}, the \textsc{epqe} takes place, giving rise to a (quantum) relativistic state transition.}
\subsection*{The Reciprocity of Heat Exchange} % \label{FluxPhotons}
According to the theorem~\pc{LewisdS}, the heat exchange between the thermal reservoir and the perfect vapor can be described as a flux of photons: when a photon is absorbed by (created in) the perfect vapor, it is correspondingly emitted by (annihilated in) the heat reservoir. Conversely, when it is emitted by the perfect vapor, it is absorbed by the thermal bath.

\tpk{eachother}{Instead of imagining a process in which an entity is created from nothing or annihilated to nothing, we depict these processes as an exchange, as occurs in a commercial transaction in which two people give something to each other at the same time.}

\RMK{It is usually assumed in spectroscopy that infrared radiation acts only on the vibration and rotation of the molecules of a substance. However, according to the theorem \pc{condens}, this hypothesis cannot explain the formation of the partial entropy \REF{Sq} of the liquid phase. The results obtained above allow conclude that the action of heat on the vapor state cause chemical reactions involving the different substances that make up the two phases of the process, whatever is the region of the radiation spectrum in which it occurs.}
\subsection*{Spacetime versus Timeless Space} %\label{NewtMink}
A full discussion of the meaning of time in physics\footnote{For a discussion on this subject, see Rovelli~\CITE{Rovelli}.}  is outside the scope of this paper. It is mentioned here, exceptionally, because this approach seems to add further confusion to the role of time played in our perception of the phenomenal world.

It is important to note that the \textit{four-dimensional} interval~$dS$ has two distinct representations: in the space-time and in the timeless Minkowskian space. While in the Minkowski space, it has the following timeless representation, 
\begin{equation}\label{RelatMink}
ds^2 = x_1^2+x_2^2+x_3^2+x_4^2,
\end{equation}
in the space-time its representation,
\begin{equation}\label{RelatEins}
ds^2= \underbrace{c^2t^2-\overbrace{\left(x_1^2+x_2^2+x_3^2\right)}^{\rm Euclidean}}_{\rm space-time}.
\end{equation}
which describes the relativistic restriction (\textit{taut constraint} $|v|\le c$), imposed on the movement of Newtonian particles.

The radical difference between the structures of the gaseous and liquid phases suggests that the equations \REF{RelatMink} and \REF{RelatEins} describe distinct realities and, therefore, are not \textit{equivalent}. While the former applies to the liquid phase, characterized by molecules, whose structure is stationary and involves a non-observable variable, the spin, which is defined in the timeless Minkowskian domain, the latter applies to the gaseous phase, which is \textit{perceived} as a system of classical particles evolving in the space-time, where they can, in principle, be observed as chemical phenomena.

The internal state of a molecule is broken during a chemical reaction, that is, it undergoes a transmutation caused by an external agent (radiation), during which the \textsc{epqe} occur.

\RMK{The complete physical meaning of the elementary process described by the ``complex energies'' \REF{+Deltap} and \REF{-Deltap} will be given later\footnote{See Section \ref{A1}.}, by quantum mechanics, when we will find that they can be derived from the \emph{ladder operators} representation of the harmonic oscillator. These operators will be shown to properly describe the \textsc{epqe} that act on the clusters that compose the liquid phase of the perfect vapor.}

It is worth noting that these equations can be recognized as consequences of kinetic theory derived from Bartoli thermodynamics, thus providing a preliminary demonstration of the following proposition:

\tpk{SpRelPrcdQM}
{In the epistemological order of its formulation, quantum theory must be preceded by special relativity.}

The confirmation of the proposition \pc{SpRelPrcdQM} requires a reformulation of the Boltzmann principle and of the axiom~\ref{frAxiomE} of quantum theory.

These considerations suggest the following conjecture:

\tpk{colapsus}{The phenomenon known as the \emph{collapse of the wave function} (which occurs with a molecule during a chemical reaction) is the change that occurs in the structure of a many-body quantum system caused by the action of the \textsc{epqe} between matter and radiation.}
\subsection*{Time as a Dependent Variable}
The elementary processes of absorption and emission correspond to the elementary (discrete) variation of the entropy $\Delta S=\frac{\Delta Q}{T}$ which occurs during a thermodynamic process{, {\sf i.e.}}, a chemical reaction. That magnitude gives the ticking of the chemical clocks that characterize each variation of the thermodynamic elementary system, as predicted by the theory of special relativity. Since all perceptions we can get from the world are given by the chemical transducers in our organisms, the world we see is only the universe of the \textit{time varying entropies}.  

The hypothesis \pc{AbsEmiss} implies a time-dependent equation, which stems from the conditional probabilities that characterize the ladder operators\footnote{The equation here mentioned is the Markov differential-difference equation that characterizes the birth and death processes, where the conditional probabilities can be treated as mutually independent agents of change~\CITE{Feller}.}, thus distinct from the interpretation of the time as an independent variable given by Schroedinger in his equation.

Recall that a wave-function $\Phi(x,t)$ is the products of two independent terms~\CITE{Pauling},
\[ \Phi(x,t)=\phi(x)\varphi(t). \]
It is known that the imaginary character of the differential term, $\imath\delp Et$, endows Schroedinger equation with time-symmetric character, hence unrealistic.

The \textit{taut constraint} $|v|\le c$, imposed by special relativity on the interaction of the molecules of the perfect vapor with heat, imparts a reformulation of the function $\varphi(t)$. 
After the evidence that special relativity has epistemological precedence over quantum mechanics, we interpret the imaginary constant, $\imath$, as of Minkowski extraction, hence the function $\varphi(t)$ is complex.

\RMK{A relativistic kinetic theory implies assume the existence of certain forces hidden behind the elementary energy conversion processes, that cause non-conservative changes on the state of motion of the particles, giving rise to the \textit{internal pressure}\footnote{The partial pressure $p_i$ acquires meaning in the language of the Physical-Chemistry as the cause of the \textit{internal work}, $dW_i=p_idV$ exerted against the internal forces of the substances~\CITE{Guerasimov}.}~$p_i$.}
\subsection*{The Spin}
We have already seen that the \textsc{epqe} are relativistic. In Section \ref{SecSR}, the existence of an unexpected imaginary form of energy $\imath c\Delta p$ was revealed together Einstein's famous scalar mass-energy conversion $\Delta\epsilon=\Delta mc^2$ equation. It can be related to the spin, which, according to the algebraic representation of the imaginary part of the complex product $uv^*=(u\cdot v)+\imath(u\times v)$ {can immediately be recognized as the expression of the \textit{angular moment} and interpreted either as a rotation or as a torsion in a Minkowski plane. Both phenomena are unobservable.
\subsection*{Comments on the Uncertainty Principle}
Based on the evidence established, mainly during the \textit{Old Quantum Theory}, especially the revelation that the energy exchanged between radiation and the quantum system is given by the Einstein-Bohr relation, we are led to conclude that the equations~\REF{+Deltap}-\REF{-Deltap} describe the \textsc{epqe}, leading to the quantum relation $\Delta p\Delta q=h$, a meaning quite distinct from that given by Heisenberg's \textit{Uncertainty Principle}.
\subsection*{Chemical Kinetics}
The interpretation of \pc{eachother}, about the consequences of the elementary processes of energy conversion in the perfect vapor, is appropriate for the analysis of the kinetics of a \textit{quantum system of many-bodies}, where creation and annihilation are different names to describe the \textsc{epqe}: the chemical analysis of the vapor state provides evidence that, during the \textsc{epqe}, concurrently chemical reactions involving the components of its phases occur.

After the revelation that the elementary energy conversion processes that occur in the perfect vapor are relativistic, we can state the purpose of the forthcoming Chapters, in terms of the following proposition:

\tpk{algoper}{To find the natural laws that rule the elementary processes that cause the changes of the \emph{concentration} of the faseous phase of the perfect vapor that conduct it towards the final state of thermodynamic equilibrium.}

%5
\section{The Quantum Taut Constraint}\label{TheTaut}

Although several meanings were given during the \textit{Old Quantum Theory} to the relation, 
\begin{equation}\label{I:ur}
\Delta p\Delta q=\Delta{\cal E}\Delta t\ge h,
\end{equation}
all of them were buried and superseded by the uncertainty principle.

In the present Section, we will resume Sommerfeld's interpretation of the relation~\REF{I:ur}, which gives it a dynamical meaning consistent with both the revelations described in Section~\ref{SecSR} and the empirical evidence obtained by the chemists, that can be summarized in the following proposition:

\tpc{SommerCampo}
{Sommerfeld's Conjecture}
{Planck's constant $h$ (quantum of action) puts a lower limit to the size $p_\ell q_\ell$ for every pair of degrees of freedom, $\left\{p_\ell, q_\ell\right\}$,
\begin{equation}\label{SommerEq}
\Delta p_\ell\Delta q_\ell = h
\end{equation}}

Conjecture \pc{SommerCampo} allows translate the relation \REF{I:ur} into the language of set theory:

\tpc{AmendSommer:1}
{Sommerfeld Redefinition of the Phase Space}
{The inequality,
\begin{equation}\label{protoineq}
\Delta p_\ell\Delta q_\ell > h,
\end{equation}
describes an open subset of the {\em classical phase space}, whose {\em boundary set} is given by the equality \REF{SommerEq}.}

\tpc{tautcnstrnt}{Taut Constraint}{The relation \REF{SommerEq} describes a \emph{taut constraint}~\CITE{Gantmacher} imposed by radiation on the motion of the radicles in the clusters of the phases of perfect vapor.}

This reinterpretation of relation \REF{SommerEq} allows for solving the proposition \pc{algoper}, where from the following corollaries derive:

\tpk{corol:1}{Under the condition \REF{protoineq}, the processes that cause changes in the values of any degree of freedom, are classical.}

\tpk{{corol:2}}{The \textit{uncertainty principle} must be interpreted, no longer as a foundational axiom of quantum mechanics, but as a \textit{symptom} that merely \textit{denounces} the existence of the \emph{taut constraint} \REF{SommerEq}, thus completing the description of the relativistic processes of absorption and emission of quanta, introduced in the Section~\ref{SecSR}.}
 
%_____________________________________________________
\subsection*{Classical Analysis of the Taut Constraint}

Planck's constant has been treated as a scalar magnitude. In quantum theory, however, it is inextricably connected to the algebra of certain complex operators that give it, in Axiom \ref{frAxiomE}, the imaginary dimension. This character, as shown in this paper, is inherited from the Minkowski representation of space displacement.

Except for Planck's constant itself, all quantities in the {relations \REF{SommerEq}-\REF{protoineq} are classical. It is nonetheless legitimate and insightful to analyze them from the Newtonian perspective. It will be shown in this Section that the equation~\REF{SommerEq} can be adopted as the expression of a \textit{stochastic taut constraint} that arises when the molecules of the gas interact with radiation. We can then state the following proposition:

\tpc{wikcotc}
{Fundamental Problem of the Kinetic Theory of the Vapor}
{To show that the \emph{taut constraints} imposed on the state of motion of the molecules of a chemical substance under the influence of an electromagnetic field, reproduce and detail the constraints imposed by the second law of thermodynamics on its behavior.}}

It is known from Analytical Mechanics that the introduction or elimination of a certain class of constraints changes the number of degrees of freedom of the phase-space of a system of particles. 

\subsection{The Forces imposed by the taut constraints}\label{qForces}
{Instead of interpreting the variation $\Delta p$ and $\Delta q$ of the values of the conjugate quantities $\{p,q\}$ in {relations \REF{SommerEq}-\REF{protoineq} as \textit{measurement uncertainties}, as prescribed by the uncertainty principle, in the forthcoming they will be treated as the \textit{changes} the momentum and position a particle of the gas undergoes during a finite time interval of length $\Delta t$ under the action of a certain type of \textit{force}, described in terms of the \textit{difference operator}~$\Delta$.}

\RMK{This assumption is justified both by its thermodynamic consequences and by their correspondences with the operators of wave mechanics, as will be shown in Section~\ref{conjugateVars}.}
	
In Newtonian mechanics the work $\Delta{\cal E}$ performed by a force $f$ is expressed by the equation, 
\begin{equation} \label{Detla}
\Delta{\cal E} = f\cdot\Delta q,
\end{equation}
where $\Delta q$ is the displacement traversed by the particle. Denoting by $\Delta p$ the change of the linear momentum of the particle, equation \REF{Detla} can be written in the form,

\begin{equation}\label{workoff}
\Delta{\cal E}=\frac{\Delta p\cdot\Delta q}{\Delta t},
\end{equation}
where $\Delta t$ is the period of time during which the momentum of the particles undergoes the variation $\Delta p$.

\tpk{Torque}{In classical mechanics the expression \REF{workoff} can be rewritten in any of the two forms,
\[ \frac{\Delta p}{\Delta t}\cdot\Delta q\quad\hbox{\rm or}\quad\Delta p\cdot\frac{\Delta q}{\Delta t}, \]
an indeterminacy that is only eliminated by special relativity, as was shown in Section~\ref{MRProblem}.}

In Newtonian mechanics the meaning of force is usually assigned to the finite quantity $\Delta p/\Delta t$, and \textit{exceptionally} to the derivative operation of Calculus,
\begin{equation}\label{eq:force}
f=\lim_{\Delta t\rightarrow 0}\frac{\Delta p}{\Delta t}.
\end{equation}

Let us rewrite \REF{workoff} in the form,

\begin{equation}\label{bi:uncA}
\Delta{\cal E}\Delta t=\Delta p\cdot\Delta q.
\end{equation}
If the derivative \REF{eq:force} exists, then,
\begin{equation}\label{eq:zeroh}
\lim_{\Delta t\rightarrow 0} \Delta{\cal E}\Delta t=
\lim_{\Delta{\cal E}\rightarrow 0} \Delta{\cal E}\Delta t=0.
\end{equation}

To examine the physical meaning of \REF{bi:uncA}, when the constraint \REF{UREBr}, is added to mechanics, we invoke Bridgman's criterion\footnote{\label{BRIDGMAN}``\textit{In dealing with physical situations, the operations that give meaning to our physical concepts should be the physical operations actually carried~out}''~\CITE{BridgmanL}.}:

\tpk{RRPhy}{According to the equality in equation~\REF{I:ur}, 
\begin{equation}\label{UREBr}
\Delta{\cal E}\Delta\tau=\Delta p_\ell\Delta q_\ell=h,
\end{equation}
the action developed by motion equals its lowest value, $h$. Thus, the operation that gives it meaning is not~\REF{eq:zeroh} but instead the quantum constraint~\REF{UREBr}.}

Newton's second law provides a general indeterminate equation that is completed with the special laws of forces that act on the motion of the bodies. When the quotients $\Delta p/\Delta t$ and $\Delta{\cal E}/\Delta q$ have, for a given motion of a body, well defined\footnote{When the constraints imposed by the quantum constraint have negligible influence on the motion of the body.} limits for $\Delta t\rightarrow 0$ and $\Delta q\rightarrow 0$, then \REF{bi:uncA} can be rewritten as a second order ordinary differential equation whose solution, given its initial conditions, completely describes that motion.
	
When, as opposed, the influence of the quantum constraint cannot be neglected, a different system of ``equations of motion'' must be considered.

While in classical mechanics we can assign to $\Delta p/\Delta t$ the Newtonian meaning of force, in the analysis of quantum phenomena, there is another agent of change on the motion of the particles of relativistic nature, as noted in \pc{Torque}. The interpretation \pc{RRPhy} justifies the following assumption:

\tpc{N:taut}{Taut Constraints and Absorptions and Emissions}{The relation \REF{UREBr} describes the \emph{taut constraint} imposed on the motion of the molecules of the gas that arise only during the \textsc{epqe} when the molecules of the gas interact with radiation, absorbing or emitting a quantum.}

\RMK{Hypothesis \pc{N:taut} allows adopt Sommerfeld's conjecture \pc{SommerCampo} as the descriptor of an inelastic collision, and thereby of a chemical reaction (see Section~{\rm\ref{NUR:Binary}}).}

Taut constraints can be treated in Analytical Mechanics according to the following suggestion~\CITE{Gantmacher}:

\begin{quote}\small\it
``The motion of a system on which unilateral constraint is imposed may be divided into portions so that in certain portions the constraint is taut and the motion occurs as if the constraint were bilateral, and in other portions, the constraint is not taut, and the movement occurs as if there were no such constraint.
In other words, in certain portions, a unilateral constraint is either replaced by a bilateral constraint or is eliminated altogether.'
\end{quote}

Hence, the motion of particles whose degrees of freedom satisfy the condition \REF{protoineq} must be treated as classical. The condition \REF{SommerEq}, differently, imposes on the particles involved, the behavior described in proposition~\pc{wikcotc}.

\subsection{Mechanical consequences of the quantum constraint}\label{non-cons}

Since according to the quantum constraint \pc{RRPhy}, the limit operation $\Delta q\rightarrow 0$ does not exist, we conclude that: 

\begin{quote}\small\it
	{The derivative
\[\lim_{\Delta q\rightarrow 0}\frac{\Delta{\cal E}}{\Delta q},\]
is undefined in the scope of quantum phenomena.}
\end{quote}

According to a well-known theorem of classical mechanics,

\tpk{nncons}{The forces that arise in the motion of particles constrained by the quantum constraint are {\em non-conservative.}}

When the quantum constraint cannot be neglected, it is not only hopeless to search for an approximation for this process in terms of some suitably chosen (perturbation) potential, but especially misleading, for it conceals the existence of the elementary energy conversion processes and their intrinsic irreversibility, hindering the quantum nature of thermodynamic phenomena.

We also assume that the quantum constraint also imply that \ldots

\RMK{\ldots if, in the treatment of any magnitude that is under the action of the quantum constraint, is described by a continuous and differentiable function, the former condition prevails over the latter.}
\noindent where from the following well-known corollary derives:

\RMK{The use of the {\em limit} operation of differential calculus must be reviewed under the quantum constraint.}

This proposition has far-reaching consequences, for, by denying the existence of convergence processes that, acting on the degrees of freedom of a single particle, give rise to well defined limit values, the quantum constraint undermine the very foundations of differential calculus, thereby subverting the whole of Analytical Mechanics.

\tpk{QH:chem2}{This condition should not be surprising, for, as already noted in \pc{DestroiDOFs} and \pc{OnBB:0}, during a chemical reaction, when the molecules of the reactants become subject to the taut constraints \REF{SommerEq}-\REF{protoineq}, their classical degrees of freedom are annihilated, losing both their classical \textit{identities}\footnote{This phenomenon, that has nothing to do with the observation or measurement processes, has led to interpret expressions \REF{SommerEq}-\REF{protoineq} as \textit{uncertainty relation} by the uncertainty principle.}, and new ones, with ``unpredictable'' values, are created with its products.}

\subsection{Inelastic collisions and the quantum of action}\label{PSxThd}
A workable description of the motion of particles in the vapor can be obtained from a reinterpretation of the description of the motion of the molecules in the gas given by Maxwell$^{\rm\ref{MaxDescr}}$.

%\tpk{MaxDescr}{During the great part of their course the molecules (of the gas) are not acted on by any sensible force and therefore move in straight lines with uniform velocity. When two molecules come \emph{within a certain distance of each other, a mutual action takes place between them} {\em(\ldots)}. Each molecule has its course changed, and starts a new path~{\CITE{Maxwell}}.}

The relations \REF{SommerEq}-\REF{protoineq} are interpreted here not as the manifestation of measurement uncertainties, but as the description of a \emph{taut constraint} which gives to Maxwell's \emph{mutual action} a different meaning, namely, that of an \emph{inelastic collision of finite duration}, in the course of which the mechanical action developed by the colliding particles is determined by the \emph{quantum-relativistic laws} of absorption and emission of one quantum of action. 

Maxwell's description \pc{MaxDescr} refers to two processes: the inertial inter-collisions translation, and an inelastic collision, corresponding, respectively, to the edges $(F)$ and vertexes $(C)$ of the polygonal line 
\[ \cdots FCF\cdots CFC\cdots \]
that describes the Brownian movement.

\subsection{Action representation of Hamilton's equations}\label{EqHamilton}

According to \pc{MaxDescr} the motion of a radicle in the gaseous phase is the alternation of two elementary motions, viz., a free path and a collision. To describe them we invoke the system of Hamilton equations, 
\begin{equation}\label{Hamilton}
\dpdz qt=\del Hpq, \qquad \dpdz pt=-\del Hqp.
\end{equation}

The description of the former, being classical and conservative, is straightforward. An inelastic collision, as opposed, is a non-conservative process requiring an \textit{ad-hoc\/} treatment.

To describe the variation of the actions $\Delta \alpha_q$ and $\Delta \alpha_p$ of the two possible outcomes of a collision, we rewrite \textit{Hamilton\/}'s equations~\REF{Hamilton} in their finite forms in the time interval of duration $\Delta t$,
\begin{equation}\label{I:p}
\begin{array}{l}
\Delta \alpha_q =
\left(\Delta H\right)_q{\Delta t} = \ \ \,\Delta q\Delta p,\\
\Delta \alpha_p =
\left(\Delta H\right)_p{\Delta t} = -\Delta p\Delta q.
\end{array}
\end{equation}

We can readily recognize in \REF{I:p} Newton's second law \textit{prototypes\/} \REF{I:ur}, and thereby assume the existence of two independent phenomena, described respectively by the variations~$\alpha_q$ and~$\alpha_p$ of the actions developed during the reaction, as already proposed in~\pc{PostCriAniq}.

\subsection{Independence of Absorptions and Emissions}\label{IndAbsEmiss}
We must then assume that the energy is not invariant, {\sf i.e.,} the Hamiltonian $H_C$ of a collision is time-dependent, say, $\Delta H_C\ne 0$ during the time interval $\Delta t$ (that corresponds to the classical expression $\partial H_C/\partial t\ne 0$), whence\footnote{This fact is more formally represented by the properties of the dynamic quantities, imposed by special relativity in equations \REF{+Deltap}-\REF{-Deltap}.}, the \textit{phenomena} described respectively by $\Delta \alpha_q$ and $\Delta \alpha_p$, are independent.

We are then led to the following corollary:

\tpc{TwoElProcs}{The two elementary energy conversion processes}{To describe the interaction of a particle with its environment (a ``\emph{quantum-relativistic reaction}'') the two equations in \REF{I:p} are \textit{uncoupled}{, \textsf{i. e.}}, they are mutually independent, corresponding to two \textit{different} processes\footnote{The assumption that there are two opposing processes was emphasized in the notation adopted in equation~\REF{I:p}, where the subscripts $q$ and $p$ (usually neglected in the representation of Hamilton's equations), were imported from the partial derivatives in~\REF{Hamilton}.} of finite duration that causes the {\em quantum\/} changes of the state of the particle.}

As in Section~\ref{SecSR}, we here assume that these two processes are identified with {absorption and emission} of a photon by a system. {Corollary \pc{TwoElProcs} is implicit in Einstein's 1917 paper}:

\tpk{EinsteinHypo}{Absorptions and emissions of radiation are two \emph{independent stochastic} processes.}

\subsection{The (Thermal) Environment as a Requisite}\label{EnvReq}
The principle of conservation of energy imposes another requirement to describe the motion of particles under the quantum constraint. According to \textsf{\pc{nncons}} we must assume the existence of an \emph{environment}\footnote{Such environment is here identified with the \textit{medium} referred to by Newton in the Definition \textsc{i} of his \textit{Principia}.} where the particles move and with which they interact, exchanging the non-conservative amounts of energy (heat) emitted or absorbed during these processes.

That the interaction of material particles with a certain kind of environment is a mechanical problem was recognized in the synthesis of the electromagnetic theory. While this theory can satisfactorily characterize the nature of the environment that is susceptible to interact with electrically charged particles and explain this interaction requiring no modification of Newtonian Mechanics, it cannot explain the behavior of the electrically charged particles involving atoms and molecules~\CITE{Pauling}. Especially, it can explain neither how the environment causes changes to the behavior of the particles of a chemical substance, nor how they are reflected in their thermodynamic properties.

Thermodynamics identifies the environment with a \textit{thermal bath} (\textit{heat reservoir}) characterized by its temperature. Although the identification of the physical properties of this environment, together with the understanding of the way it interacts with material particles, are essential for the complete formulation of kinetic theory, the revelation of the mechanical nature of heat was not considered a problem of thermodynamics before Bartoli's amendment to thermodynamics.

%6
\section{The Ladder Operators}\label{TheQuantum}

{It is noteworthy that any process that causes the change of the value of the action $\alpha_\ell=\Delta p_\ell\Delta q_\ell$ of the degree of freedom $\ell$ of a system (or its very elimination), causes the same variation of the volume of the phase space of that system, and thereby, according to the Boltzmann Principle, of its entropy.}

\subsection{Taut Constraints and Entropy Formation}
According to Bohr's second postulate, the variation of the energy $\epsilon$ of a quantum system that is being moved from the state $m$ to the state $n$, is caused by one of the \textsc{epqe}, expressed by the formula,
 
\begin{equation}\label{EBr}
\Delta\epsilon_{mn}=\epsilon_m-\epsilon_n=h\nu_{mn},
\end{equation}
where $\epsilon_m$ and $\epsilon_n$ are the energy values in the states, $\{m,n\}$ under consideration. The abbreviated form of \REF{EBr}, $\Delta\epsilon=h\nu$, is known as the {Einstein-Bohr relation}.

By rewriting equation \REF{EBr} in terms of the \textit{period} $\Delta\tau=1/\nu$ of the frequency~$\nu$, we recover the well-known representation of the action, expressed in terms of the variation $\Delta{\cal E}$ of the energy during the duration $\Delta\tau$ of the process,
\begin{equation} \label{EBr-tau}
\Delta{\cal E}\Delta\tau = h.
\end{equation}

\paragraph{Reinterpretation of Heat.}
Comparing the amount of energy given by the equations \REF{+Deltap} and \REF{-Deltap}, with the corresponding value given by the equation \REF{EBr}, we conclude that the notion assigned to the amount of \textit{heat} exchanged between matter and radiation, should be revised.

In order to improve the connection between thermodynamics and quantum theory, it is imperative to remember that Carnot's \textit{working substance} is a \textit{chemical substance}{, \textsf{i. e.}}, its corpuscles are neither Newtonian particles nor Planck resonators, but atoms and molecules.

\subsection*{Algebraic Approach to the Planck Resonator}
Let us momentarily admit that, according to classical mechanics, the unit of action $h$ is a \textit{scalar} quantity. Justified by this provisional hypothesis, we can apply the rules of ordinary algebra to combine relations \REF{SommerEq}-\REF{protoineq}, thus obtaining, the equation~\REF{UREBr}.

We can rewrite \REF{protoineq} in the form,
\begin{equation}\label{ineq}
\Delta{\cal E}\Delta\tau=\Delta p_\ell \Delta q_\ell > h,
\end{equation}

\RMK{While the character of the equation \REF{SommerEq} is corpuscular (expressed in terms of degrees of freedom of translation), the equation \REF{EBr}, in opposition, is undulatory (expressed in terms of oscillatory degrees-of-freedom).}

\subsection*{The Quantum Exchange Algebra}\label{dEcomh}
Assuming that $\Delta q_\ell=\lambda_\ell$ in \REF{UREBr}, where $\lambda_\ell$ is the wavelength of the radiation, we have $\Delta p_\ell=h/\lambda_\ell$. Substituting this result in equations \REF{+Deltap} and \REF{-Deltap}, we obtain the expression of the quantum-relativistic equations that describe the \textsc{epqe} by atoms and molecules,
\begin{eqnarray}
&\Delta\epsilon_\ell   &= \Delta m_\ell c^2+\imath c\frac h{\lambda_\ell},\label{+lambda}\\
&\Delta\epsilon_\ell^* &= \Delta m_\ell c^2-\imath c\frac h{\lambda_\ell}.\label{-lambda}
\end{eqnarray}

\RMK{The equations \REF{+lambda} and \REF{-lambda} will be justified in Section~\ref{A1}, when the forces that act on the corpuscles (the radicles) of the perfect vapor are caused by the \textsc{epqe} which, in their turn, are described by the ladder operators algebra.}

From the \textit{Principle of Conservation of Action} \ref{CvtvAct} the following principle derives:

\tpc{xgQuantum}{Principle of Reciprocity}{The exchange of quanta between matter and radiation are reciprocal processes: to the \textit{absorption} of a photon by a particle, there corresponds the \textit{emission} of a photon by radiation, and \textsf{vice versa}.}
		
The three conservative principles of classical mechanics, namely the invariance of energy, of linear momentum, and of angular momentum, correspond to the invariant entities in the phase transition processes in the perfect vapor. In the absorption and emission processes, the energies involved are given by the relativistic equations \REF{+Deltap}-\REF{-Deltap}, from which the equations \REF{+lambda}-\REF{-lambda} derive. These equations establish the functional correlation involving the degrees of freedom that describe these interaction processes, namely mass, momentum and \textit{spin}, a discrete version of angular momentum.

% Preprint.tex [167]

\subsection*{The Algebra of Quantum Complex Numbers}\label{AlQComplex}
Complex numbers are usually treated as a \emph{field}. Thus, for each complex number $u\ne 0$, there is its inverse $ u^{-1}$. This is not true in the algebra of the quantities \REF{+Deltap}-\REF{-Deltap}, because the complex conjugate $u^*$, required to calculate $u^{-1}$, corresponds to a physical process that is \emph{independent} of~$u$.

This evidence requires a reinterpretation of the Axiom~\ref{frAxiomC}, which provides the instructions to conceive the algorithm used to \emph{artificially} calculate the eigenvalues of a given dynamic variable. In the formulation of such \textit{in silico} algoritm, no consideration about what is the \textit{in vivo} algorithm, effectively executed by the real particles, which is the result of the reiteration of stochastic processes of absorption and emission of quanta, described by their corresponding ladder operators which are \textit{mutually independent}.

%__6________________________
\subsection{Chemical Reactions and Inelastic Collisions}\label{ChReactInelColl}\label{NUR:Binary}
To reconcile the variation of the energy given by \REF{EBr} with the relativistic variations described by equations \REF{+Deltap} and \REF{-Deltap}, we have to inquire, not about the meaning of Planck's constant, itself, but about the algebraic structure it imposes on the mathematical representation of these phenomena, which leads us to inquire on the laws that rule the chemical reactions and their quantum representation.

The chemical reaction is interpreted by kinetic theorists as an inelastic collision. In classical mechanics, a binary collision is generally treated as a two-dimensional motion. Taking the origin of coordinates at the center of mass of the two colliding particles, the Lagrange function of the motion in the plane formed by the relative initial and final velocities, can be expressed in terms of complex variables,
\begin{equation} \label{LM:14,1}
{\cal L} = \frac m2\left(\Delta\dot z\right)\left(\Delta\dot z^*\right),
\end{equation}
where $\Delta z$ is the vector describing the trajectory of the system, and $\Delta z^*$ is its conjugate.

During the evolution of the collision process, a certain amount of action $\Delta\alpha={\cal L}\Delta t$ is produced. In its evaluation, an indeterminacy arises, because the pre-multiplication and post-multiplication of $\cal L$ by $\Delta t$ give different results. Since, for the purposes of this present approach, the inelastic collisions we are dealing with here is the same phenomenon already discussed in the Section~\ref{SecSR}, we must acknowledge the hypothesis \pc{EinsteinHypo}, which states that the actions $\Delta\alpha$ and $\Delta\alpha^*$ refer to two different physical variables. Hence, 
\begin{equation} \label{L2N:nwith}
\left(\Delta t\frac{\Delta z}{\Delta t}\right)\left(m\frac{\Delta z^*}{\Delta t}\right)
\ne
\left(m\frac{\Delta z}{\Delta t}\right)\left(\frac{\Delta z^*}{\Delta t}\Delta t\right),
\nonumber
\end{equation}

The hypothesis \pc{EinsteinHypo} will then be superseded by the following postulate:
	
\tpk{ThereAre}{There are in nature two independent phenomena, characterized respectively by the action $\Delta\alpha$ 
and~$\Delta\alpha^*$, which correspond, respectively, to the \textsc{epqe}.}

Postulate \pc{ThereAre} implies rewrite the members of \REF{L2N:nwith} separately,
\begin{equation}\label{Ldt}
\begin{array} {ll}
\Delta\alpha & = {\cal L} \Delta t = %M\frac{\Delta z}{\Delta T}\Delta Z^ * =
(\Delta p)\left(\Delta z^*\right),\\
\Delta\alpha^* & = \Delta t {\cal L} = %m \Delta z\frac {\Delta z^*} {\Delta t}=
\left (\Delta z\right) \left(\Delta p^*\right).
\end{array}
\end{equation}

\subsection{Bose Operators}\label{A1}
Rewriting actions \REF{Ldt} in dimensionless form, we have,
\begin{eqnarray}
\frac{\Delta\alpha}h   &=&\frac{\Delta p\cdot\Delta z}{h}+\imath\frac{\Delta p\times\Delta z}{h},\label{CA:ops1}\\
\frac{\Delta\alpha^*}h &=&\frac{\Delta p\cdot\Delta z}{h}-\imath\frac{\Delta p\times\Delta z}{h}.\label{CA:ops2}
\end{eqnarray}
where from we obtain,
\begin{equation}\label{DADA*}
\left(\Delta\alpha-\Delta\alpha^*\right)=2\imath(\Delta p\times\Delta z).
\end{equation}

Identity \REF{DADA*} allows recognize, in $\Delta\alpha/h$, and $\Delta\alpha^*/h$, the elements of the complex algebra representation of the \emph{quantum creation and annihilation operators}, $\left\{{\sf a}^\dag,{\sf a}\right\}$, which characterize the Bose-Einstein statistics, that can be also derived, for instance, from the Planck resonator (harmonic oscillator). The angular momentum representation derives immediately from the algebraic meaning of the imaginary part of the complex product, as can be immediately seen in equations \REF{CA:ops1} and \REF{CA:ops2}.

\subsection*{The Second Law and the Quantum Constraints}\label{2LawConstraints}
The persistent action of absorption and emission of quanta causes the \textit{shuffling} of the particles, giving rise to collective, macroscopic behavior, eventually comprehended as the constraint imposed by the Second Law of Thermodynamics.

As known from Analytical Mechanics, the addition to (or removal from) a system of particles, of a certain class of \textit{constraints}, cause a change in the number of its degrees of freedom and therefore, according to the Boltzmann Principle, of its entropy. It is then possible to consider a chemical reaction as a consequence of the \textit{quantum-relativistic} constraints which impose stochastic restrictions on the motion of the molecules.

\subsection{Taut Constraints and the Schroedinger Equation}\label{conjugateVars}
Chemistry reveals that a chemical reaction causes the following additional consequence of fundamental importance for quantum theory:

\tpc{DestroiDOFs}{Destruction and Creation of Degrees of Freedom}{During a chemical reaction the degrees of freedom of the reactants are \emph{destroyed}, and the resulting products that are \emph{created} acquire new degrees of freedom with unpredictable values.}

The evidence \pc{DestroiDOFs} leads us to inquire on the following question:

\tpk{ChR-XrEq}{How the independent degrees of freedom of a radicle in the gaseous phase of the perfect vapor become, as the consequence of a chemical reaction, \textit{conjugate} inside a cluster?} 

Let us derive from equation \REF{UREBr} the finite work performed by the force,
\begin{equation}\label{qWork}
\Delta{\cal E}=\frac {\Delta p_\ell}{\Delta\tau}\Delta q_\ell = h\nu.
\end{equation}

If this algebraic manipulation is justified, we are allowed to adopt the equation \REF{UREBr} as the expression of a non-conservative process.

The quantity $\frac{\Delta p_\ell}{\Delta\tau}\cdot\Delta q_\ell$ in the equation \REF{qWork} can be interpreted as the work exerted either by the radiation on some degree of freedom of the material medium during the absorption of a quantum or the reverse during its emission.

The connection between the pairs of degrees of freedom $\{p_\ell,q_\ell\}$ or between $\{{\cal E},\tau\}$, can be justified when we realize that Planck's constant cannot be interpreted as a mere scalar magnitude. Due to the enigmatic nature of Planck's constant, the mathematical interconnection of these pairs of quantities were given in the formulation of wave mechanics by an unconventional treatment of the involved mathematical terms, as follows.

Rewriting equations \REF{SommerEq} and \REF{EBr} in the forms,
\begin{equation} \label{eq:delx}
\Delta {\cal E} = \frac h {\Delta t},\qquad\hbox{and}\qquad
\Delta p=\frac h {\Delta q},
\end{equation}
the difference operators $\Delta$ become interpreted, in quantum mechanics, as the discrete version of the corresponding differential operators of energy and momentum, respectively,

\begin{equation}\label{EpA}
\lim_{\Delta t\rightarrow 0}\frac h{\Delta t}\rightarrow\hat{\cal E}=\frac h{2\pi\imath}\frac\partial{\partial t},
\qquad\hbox{\rm and}\qquad
\lim_{\Delta x\rightarrow 0}\frac h{\Delta x}\rightarrow\hat p_x=\frac h{2\pi\imath}\frac\partial{\partial x}.
\end{equation}

The correspondences \REF{EpA} and the algebraic properties of operators adopted in quantum mechanics establish the rules necessary to formulate the operators that characterize atoms in atomic theory and molecules in chemistry, as many-body quantum systems. In an equation, these operators can be interpreted as \textit{conjugators}{, \textsf{i. e.}}, entities that \textit{merge} the degrees of freedom, originally independent, into a single \textit{conjugated} unit, imposed on them by the quantum exchange processes.

Besides, the Planck constant $h$ appears in these operators always combined by algebraic rules with the imaginary unit $\imath$ so that we can treat their combination as a single imaginary entity $\imath h$.

Schroedinger's equation can be interpreted as a translation of the characteristic Hamiltonian of a mechanical system of particles into the language of operators\footnote{This procedure can be justified by Bohr's \textit{Correspondence Principle}.} It is here hoped that a more reliable derivation of the limit operation \REF{EpA} from the \REF{SommerEq}-\REF{protoineq} \textit{taut constraints}, will be achieved by mathematicians in the near future. This formulation would authorize us to assign to these relations the \textit{operative faculties} that have stayed hidden in previous interpretations.

\RMK{By the \emph{operative faculty} of an agent we understand its "power to cause a change" in the entity in which it acts. Although the notion of force is criticized by some authors~\CITE{Jammer}, the mathematicians borrowed its meaning from Newtonian physics to establish how the notion of an operator should be interpreted.}

From the previous considerations we conclude:

\RMK{The motion of the gas molecules in the phase sub-space \REF{ineq} is classical, while the identity \REF{UREBr} refers to a finite elementary process whose duration is given by the period~$1/\nu$, during which the volume of the phase space of the perfect vapor, and thereby its entropy, undergoes a change.

Furthermore, by having recourse to the results obtained in Section~\ref{SecSR}, and acknowledging the algebraic imaginary operating character it imparts to the differential coefficients of Schroedinger's differential equation, we can finally endow a discrete fluid with susceptibility to heat.}

\subsection{The Causes of Change of the Phase Space Volume}\label{PostCriAniq}
Equations \REF{CA:ops1} and \REF{CA:ops2} describe the changes that the dimensionless volume of the phase-space of the perfect vapor undergo under the action of radiation, leading us to conclude:

\tpk{Ps=Mink}{The algebraic structure of \emph{phase space} of a system of particles interacting with radiation is Minkowskian.}

The revelation \pc{Ps=Mink}, which eliminates the \textit{ambiguity} already detected in Section~\ref{conjugateVars}, implies the need to reformulate the Boltzmann principle, by adopting $\theta$ as the variable that endows its functions with thermodynamic and relativistic faculties.

Introducing the definition,
\begin{equation}\label{Caca}
\Delta p\times\Delta z
= - \imath\frac h2.
\end{equation}
where $h$ is Planck constant, we recover, in $\Delta\alpha/h$, and $\Delta\alpha^*/h$, the {\em complex number representation} of the creation and annihilation dimensionless operators $\{{\sf a}^\dag,{\sf a}\}$ that characterize the statistical indistinguishability of the particles in the thermodynamic equilibrium of a gas of Bose particles (the Bose-Einstein statistics).

\RMK{Note that the definition \REF{Caca} gives physical (\emph{relativistic}) meaning to the Axiom~\ref{frAxiomE}.}

\subsection{The Thermodynamic Equilibrium}\label{TheEquil}
\begin{flushright}
\textit{Eadem mutata resurgo.}

Epitaph to Jakob Bernoulli I
\end{flushright}
In quantum theory the \emph{measured value} of a dynamic variable is given by the eigenvalue of its associated operator. However, in the Second Quantization, the dynamic variables are the populations of the quantum states, whose kinetics is determined not by Schroedinger's time equation, but by the space alone equation, which is an \textit{automorphism}.

It is well known from thermodynamics that no meaning can be assigned to the entropy of the gas unless every transient internal dynamic process has been consummated. Hence, while large scale variations of these numbers are still occurring, one cannot speak of average values, and no meaning can be assigned to its wave function. Correspondingly, the wave function of a many-body system, that ultimately specifies the population of its quantum states, cannot be adopted to describe this system if it has been removed from equilibrium.

To decide between these two opposing hypothesis, viz., the statistical equilibrium versus the necessary conditions imposed by the symmetric or anti-symmetric character of the wave function, we need to recognize that the wave function of a system of particles, as much as it happens with entropy, is defined only in the \textit{thermodynamic} equilibrium state. Since indistinguishability has been interpreted as an ontological condition, we must reconsider the scope of these functions, especially of a system of Bose particles, to agree with the equilibrium of Planck's radiation law.

Recall that wave functions are the way the Schroedinger equation specifies those stationary states predicted by Bohr in his first postulate. In Schroedinger's equation, however, stationary states are not postulated, but are, instead, consequences of its definition as the \textit{automorphism}, $\hat H\psi=E\psi$.

From its algebraic definition, an automorphism can be adopted as a rigorous definition of equilibrium. As in Jakob Bernoulli's epitaph, that properly describes the logarithmic spiral curve which, after transformed (undergoing its \textit{characteristic} rotation and homothety), results in itself, any system defined by an automorphism, after undergoing its \textit{characteristic transformation}, continues in the same state it was before the change.

Each quantum system has its own \textit{characteristic} transformation, expressed in terms of its \textit{Hamiltonian operator}, of which the equilibrium state gives physical meaning to its eigen-states\footnote{If the system is already in the equilibrium state, after undergoing its characteristic Hamiltonian transformation, it remains there.}.

We are then suggested to assume that ``\textit{to be} indistinguishable'' is not an ontological character of the particles of a gas, but, instead, the result of a certain general faculty the gas is endowed with, namely, that of \textit{becoming}, under certain circumstances, susceptible to radiation. It is the persistent interaction between the gas and radiation that leads the particles to \textit{become} indistinguishable in its final state of equilibrium.

These considerations lead us to reinterpret Axiom~\ref{frAxiomC}, for the automorphism defines a \textit{condition of equilibrium}, one can give a new meaning to the same equation, by maintaining the same premise. We can rewrite the statement above in the following terms:

\RMK{Les valeurs de la variable dynamique donn\'e dans l'\'etat d\'efini du syt\`eme dans l'équilibre thermodynamique sont les valeurs propres de l'op\'erateur associ\'e $F$.}

While a variety of algorithms to derive the eigenvectors of an operator and their corresponding eigenvalues can be conceived by mathematicians, physicists are required to answer the following question:

\tpk{What}{What is the ``algorithm'' performed, not by an artificial \emph{in silico} process, but by a natural system moving towards its equilibrium state?}

From fluctuation phenomena revealed by thermodynamics we conclude that the quantum state occupancy numbers are \textit{random numbers}, so that their values vary according to the \textit{persistent stochastic iteration of the same invariable agents} throughout the relaxation time, until the eigenstate (equilibrium value) is achieved{, \textsf{i.e.}}, until each state contains the equilibrium \textit{random number} of particles given by the automorphism. An appropriate description of this kinetic process is obtained by the Markovian birth-and-death stochastic process, as described in~\CITE{CZM}.

The asymptotic value of the solution of the Markovian birth and death equation for $t\rightarrow\infty$ describes the equilibrium condition, whose eigenvalues give, not the \emph{measured} values of a dynamic variable at a given instant, but its values at the thermodynamic equilibrium (stable and \textit{supposedly}\footnote{Depending on further inquiries on this issue.} metastable).

\subsection*{On Being and Becoming}

From \pc{DestroiDOFs} derives the following corollary:

\tpk{OnBB:0}{The classical degrees of freedom, the momentum, and position of a particle of the perfect vapor, are originally independent in the gaseous phase. When the particle, after being subjected to a non-conservative process, which occurs during a chemical reaction, is transferred to a cluster in the liquid phase, their degrees of freedom thenceforward, become dependent, as described by the \emph{differential conjugate operators} that link them within this phase.}
	
Corollary \pc{OnBB:0} enforces the understanding that ``\emph{conjugation}'' is a previously non-existent \emph{relation} between momentum and position, expressed by the Hamiltonian combination of the operators \REF{EpA} in the composition of Schroedinger's equation. Such relation can be ``\emph{created}'' (come to be) or ``\emph{destroyed}'' (pass away) during an \textsc{epqe}.

The previous discussion enforces the understanding that ``\emph{conjugation}'' is a previously non-existent \emph{relation} between momentum and position, expressed by the Hamiltonian combination of the operators \REF{EpA} in the composition established by Schroedinger equation.

According to chemistry and to its consequence to quantum theory, we cannot say that position and momentum are ``something'' that \textit{always existed together} with the particles (the molecules), a conclusion that raises a challenge to the interpretation of the uncertainty principle, and a serious objection to its acceptance as a fundamental principle of molecular mechanics.

\RMK{The definitions of position and momentum, as fundamental degrees of freedom in classical mechanics, acquire meaning only if referred to the same instant of time, a notion that has a characteristic meaning in special relativity, expressed in terms of the notions of \textit{simultaneity} and \textit{space-time interval}. Note that relativistic simultaneity (zero length of space-time interval\footnote{This condition might be related with the notion of \textit{entanglement}.}) is a precondition for the consummation of the phenomenon ``\emph{to become conjugate}'', as discussed in Section~\ref{conjugateVars}.}

The \emph{Principle of Reciprocity}~\pc{xgQuantum} can then be naturally extended to the phenomenon of condensation: 

\tpk{RecpHypo}{In the perfect vapor, to the annihilation of one radicle in the liquid phase corresponds its creation in the gaseous phase, and \textsf{vice versa}.}

In the present approach, the meanings of the words \textit{creation} and \textit{annihilation} are given in~\pc{eachother}. With this understanding, statement~\pc{RecpHypo} describes a transaction, in which the words \emph{creation} and \emph{annihilation} acquire the complementary meanings of ``death and birth (reincarnation)'', as metaphors for the reciprocal processes that correspond to the exchange of a radicle between two coexisting phases. This reinterpretation justifies having recourse to the Markovian theory of birth and death processes where from the kinetics of the population of the clusters in the liquid phase derives.

\subsection*{Taut Constraints and Entropy}

The quantity \REF{SommerEq} describes the variation of volume $\Delta\alpha$ of the phase space of the gaseous phase of the perfect vapor. According to the Boltzmann Principle, such quantity corresponds to the Sackur-Tetrode entropy. 

The changes on the populations of the quantum states of the clusters in the liquid phase of the perfect vapor, are caused by the \textsc{epqe}, induced by the taut constraints $|v|\le c$ and $\Delta p\Delta q\ge h$ (here represented by definition \REF{Caca}), imposed, respectively, by special relativity and the quantum condition.

During these processes, known as \textit{absorption or emission of quanta}, a certain quantity of \textit{action} is exchanged between the substance and radiation. Concurrently radicles are exchanged between the gaseous and the liquid phases of the perfect vapor. The relativistic variation of the momentum $\Delta p=\epsilon/c$ is known to correspond to the quantum expression, $\Delta p=h/\lambda$.

Henceforth, in the description of the phenomena involved in the perfect vapor, the two taut constraints are combined into a single representation, expressed in terms of non-commutative complex quantities which, in formal quantum mechanics are represented by the ladder operators\footnote{See \S 41~\CITE{Chpolski}} $\left\{\mathsf{a},\mathsf{a}^\dagger\right\}$. 

%7
\section{Entropy Formation Mechanisms}\label{EntropyChng}

\begin{quote}
In this Chapter, the particles in Brownian Movement are considered to be in persistent interaction with a heat reservoir.

This approach extends Boltzmann's entropy to properly reflect the relativistic nature of the \textsc{epqe}, ruled by the bosonic \textit{laws of change} determined by the equations \REF{CA:ops1} and \REF{CA:ops2}. These equations can be also identified with the variation of the volume of the phase space, thus revealing the entropy formation mechanism.
\end{quote}

\subsection{The Bose Character of Radiation}\label{ActInvariance}
It can be easily verified that the equations \REF{CA:ops1} and \REF{CA:ops2} are the \emph{complex algebra representatives} of the creation and annihilation dimensionless operators $\{{\sf a}^\dag,{\sf a}\}$, derived either from the harmonic oscillator or from the quantization of the angular moment, that characterize the Bose-Einstein statistics. In this paper they are, differently, derived from the equations \REF{+Deltap} and \REF{-Deltap}.

\subsection{The Interaction of Matter with Radiation}
To properly represent the action developed during the \textsf{epqe} (which describe the increase or decrease of the dimension of the phase space by one quantum unit), we will adopt the complex equations \REF{CA:ops1} and \REF{CA:ops2} for the inner and outer products in a single mathematical expression. It provides a clear differentiation between the variation of the action during these processes from its corresponding classical representation, expressed by inequality \REF{ineq}.

The volume of the phase space of the gas is then revealed to be represented by a complex number, thus extending the definition of Boltzmann's entropy. For the particles moving in the classical subspace of the phase space, $\Delta\alpha_j>h$, the entropy is given~by
\[
\Delta \mathsf{S}_j=\ln\left(\Delta p_j\cdot\Delta q_j\right).
\]

Denoting by $\Delta\alpha\left({\mathrm{a}}^\dag\right)$ and $\Delta\alpha\left({\mathrm{a}}\right)$ the variation of the action, generated by the operators \REF{CA:ops1} and \REF{CA:ops2}. We can write the variation of the entropy by substituting the scalar by the complex logarithmic function ({\tt Ln}) in the Boltzmann principle%
\footnote{In the following expressions the superscripts $c$ e $q$ designate the classic e quantum behaviors, respectively. The signals of the equations \REF{Act:S2} depend on the system to which they refer, if radiation or the gas.}, when the gas absorbs or emits one quantum,% for $\Delta\alpha_k=h$,
\[
\Delta\mathsf{S}_k=
\left\{\begin{array}{ll}
{\rm Ln}\left[\Delta\alpha_k\left({\mathrm{a}}^\dag\right)\right]&={\rm Ln}\left[\left(\Delta p_k\cdot\Delta  q_k\right)-\imath\left(\Delta p_k\times\Delta q_k\right)\right],\\
{\rm Ln}\left[\Delta\alpha_k({\mathrm{a}})\right]&={\rm Ln}\left[\left(\Delta p_k\cdot\Delta  q_k\right)+\imath\left(\Delta p_k\times\Delta  q_k\right)\right].
\end{array}\right.
\]

{In summary,\small
\begin{eqnarray}
\Delta\mathsf{S}_j&=&\ln\left(\Delta p_j\cdot\Delta q_j\right),\qquad\hbox{\rm for}\quad\left(\Delta\alpha_j>h\right),\label{Act:S1}\\
\Delta\mathsf{S}_k&=&\left\{\begin{array}{ll}
{\rm Ln}\left[\Delta\alpha_k\left({\rm a}^\dag\right)\right]&={\rm Ln}\left[\left(\Delta p_k\cdot\Delta  q_k\right)-\imath\left(\Delta p_k\times\Delta q_k\right)\right]\\
{\rm Ln}\left[\Delta\alpha_k({\rm a})\right]&={\rm Ln}\left[\left(\Delta p_k\cdot\Delta  q_k\right)+\imath\left(\Delta p_k\times\Delta  q_k\right)\right]
\end{array}\right\}\ \hbox{\rm for\ }\ \left(\Delta\alpha_k=h\right)\label{Act:S2}.
\end{eqnarray}
}

\subsection{The Equilibrium Formation Processes}\label{EqFrmProcs}
Since the \textsc{epqe} involve certain forces that cause the changes on the motions of the particles of the gas, we have some reason to believe that they are the main responsible for the processes of equilibrium formation in the perfect vapor. An amendment to Boltzmann principle, referring to the equilibrium condition, expressed in terms of the balance between {absorptions and emissions of quanta} by the particles of the perfect vapor, is here proposed.

The \textsc{epqe} are responsible for the variation of the \textit{dimension} of the phase space. Since they are stochastic and independent, they cause a fluctuation of the volume of the phase space, that corresponds to the variation of the entropy of the perfect vapor.

\rmk{Neutro}{The change of the entropy caused by the ``{\it combination\/}'' of the two opposite {processes of absorption and emission}, is given by the amount,
{\small
\begin{equation}\label{Act:E-A}
{\rm Ln}\left({\rm a}_k^\dag\right)-{\rm Ln}\left({\rm a}_j\right)
={\rm Ln}\left[\frac
{\left(p_k\cdot q_k\right)-\imath\left(p_k\times q_k\right)}	
{\left(p_j\cdot q_j\right)+\imath\left(p_j\times q_j\right)}\right].
\end{equation}}}

Denoting the quotient,
\begin{equation}\label{FGC:vartheta}
\vartheta=\frac{p\times q}{p\cdot q}
\end{equation}
we can rewrite equation \REF{Act:E-A} in the dimensionless form,

\begin{equation}\label{Act:E/AX}
\left(
\frac{a_k^\dag}{a_j}\right)=
\frac
{1-\imath\vartheta_k}
{1+\imath\vartheta_j}.
\end{equation}

Since the entropy of the perfect vapor is given by \REF{eos:Phip}, equation \REF{Act:E/AX} can be rewritten in terms of the single magnitude~$\theta$:

\RMK{Both the quantities $\vartheta_k$ and $\vartheta_j$ are described by the same \textsc{pdf}. We can then replace the quotient $\left(1-\vartheta_k\right)/\left(\vartheta_j\right)$
by a function $g(\theta)$ of a single argument $\theta$, and rewrite \REF{Act:E/AX} in the form,
\begin{equation}\label{lnAct:E/A1}
S=\imath{\rm Ln}\left(
\frac{\avg{a^\dag}}{\avg a}\right)
=2\arctan\left(g(\theta)\right).
\end{equation}}

Equation \REF{lnAct:E/A1} is insufficient to determine the entropy of the perfect vapor. As pointed out in~Section~\ref{WienK}, the functional form of $g(\theta)$ depends on the matching of two frequency spectra, namely, of radiation and of the characteristic of the chemical substance. Hence, we cannot predict the thermodynamic properties of the perfect vapor, unless the frequency spectrum of the radiation with which it interacts is given. To properly characterize its thermodynamic properties, it is convenient to establish, by \textit{convention}, a \textit{standard radiation} spectrum, an issue that is the subject of the Section~\ref{theBlackBodyRad}.

\subsection{Potential Theory of the Equilibrium of the Liquid Phase}
According to Krönig's description of the motion of a particle in the gas, the trajectory of a particle can be described by the polygonal line,
\[ \ldots \textsc{fcfcf} \ldots, \]
where \textsc{f} denotes the intertial displacement of length $\Delta q$ traversed by the particle between two consecutive collisions and the edges \textsc{c} denote elastic collisions.

The space \textsc{f} can be represented by a free (slide) vector, as in Perrin's analysis of the Brownian movement~\CITE{Perrin}:

\RMK{Une autre vérification plus frappante encore, dont je dois l'idée à \emph{Langevin}, consiste à transporter parallèlement à eux-mêmes, les dé\-pla\-ce\-ments horizontaux observés, de façon à leur donner une origine commune (\ldots) Cela revient à considérer des grains qui auraient même point de départ.}

The non conservative character of the collisions \textsf{c}, we review Krönig's descrition so that, after sliding to the point $+\imath$ of the complex plane the origins of the free vectors corresponding to the actions developed by a particle immediately after the absorption of a quantum from radiation; and sliding to the point $-\imath$ of such plane the endpoints of the free vectors immediately before the emission of one quantum. 

Rewriting the equations \REF{+Deltap} and \REF{-Deltap} in the dimensionless form,
\begin{eqnarray}
\frac{\Delta\alpha}h  &=&\frac{\Delta mc}{h}+\imath,\label{CA:1}\\
\frac{\Delta\alpha^*}h&=&\frac{\Delta mc}{h}-\imath,\label{CA:2}
\end{eqnarray}
we can recognize in equation \REF{Act:E-A} the representation of the flux of quanta in the perfect vapor, with a \textit{source,} in the point $+\imath$ and a \textit{sink} in the point $-\imath$ of the complex plane\footnote{The description of such flux can be seen in fig. {111, Ch \textsc{iii} \S 2, p. 252}~\CITE{Lavrentiev}.}.

%8
\section{Principle of Frequency Match}\label{theBlackBodyRad}

\subsection{Frequency Spectrum of a Discrete Fluid}

A chemical bond linking the atoms in a molecule can be roughly likened to a kinematic pair %~\CITE{Reuleaux}
connecting two physical parts that impose a constraint on their relative movement, allowing regard the molecule as a \textit{mechanism}. This analogy allowed the kinetic theory to assume that parts of a crystal can oscillate (Einstein (1907), Debye (1912)) and molecules in a gas (Bjerrum (1911-1914)) can store vibrational and rotational energies thus contributing to the specific heat of substances~\CITE{Kragh}. The set of these degrees of freedom form the frequency spectrum characteristic of the substance in thermodynamic equilibrium. 

\ifodd 52
Kirchhoff's approach to spectroscopy revealed that the characteristic that differentiates one chemical element from the others in the periodic table is its frequency spectrum, which was later shown to be determined by the set of its quantum states. He predicted that, for the particular case of the Black Body Radiation, the functional form of this radiation is universal, whose formula was later determined by Planck.

Similarly, the thermodynamic state of a chemical substance also depends on the \textit{characteristic frequency spectrum} of the radiation under which it is subjected. To explain this phenomenon, molecules were compared to mechanisms in kinetic theory. The rotations and vibrations of parts of its atoms are imagined as constrained by the different articulation points within them. The independent movement of these parts has been the main source for predicting their frequency spectra. Nevertheless, as stated in~\pc{ThdBP:Potn}, the properties of perfect vapor depend also on the magnitude $\theta$ which cannot be explained in terms of classical mechanisms.

%\TB{O espectro de moléculas é discreto porque os átomos dentro delas é limitado.

%\href{https://en.wikipedia.org/wiki/Continuous_spectrum}{Continuous spectrum}}

There is yet no theoretical approach that allows predicting all the geometrical arrangements the radicles can acquire in the clusters of the liquid phase. The experimental determination of the frequency spectra of liquid water has been shown to be a difficult task~\CITE{WikiWater}.

According to its definition \pc{tVap}, there is an indeterminate number of discrete fluids for the same argument $u_{hc}$ of the entropy $F(W)$,
\[ u_{hc}=\frac{h\nu}{kT}=\frac{hc}{\lambda kT}\propto \frac{hc}{kT}\sqrt[3]{\frac NV}. \]

This function varies with the value of the frequencies $\nu$ which are supposedly continuous quantities, so that the set of these values is usually referred to by the \textit{frequency spectrum of the radiation}.
\fi %2

\subsection{Mathematical Representation of Spectra}\label{GF}
To describe discrete frequency spectra in statistical mechanics it is customary to use the \emph{generating function}~\CITE{Feller} ({\sc GF}) of the operational calculus as the mathematical method. Considering that the equilibrium formation between radiation and the chemical substance is a stochastic process, we adopt the random integer occupancy numbers~$\rj$ of the quantum states as independent coordinates and, as the dependent variable, the corresponding probability Prob$\left(\rj=k\right)$ that the occupancy number acquires the value~$k$.

%The polynomial series with probabilities as coefficients (the \emph{partition function})\footnote{See Section \ref{Ctes}, where we showed that the argument of the characteristic that the entropy of the radiation is $\Ucte=hc$, so that $s=\frac{h\nu}{2kT}$.} has been used to describe the frequency spectra in terms of the {\sc pdf} of the particles in the Fermi and Bose gases.

The {\sc gf} of the \textsc{pdf} of the occupancy random integer variable ${\bf r}_\ell$ is well known from the description of radiation in terms of the quantum harmonic oscillator,
\begin{equation}\label{csch}
\beta(s)=\frac 12\ {\rm csch}\left(\frac{h\nu}{2kT}\right).
\end{equation}

To properly describe a thermodynamic system it is necessary to know the spectrum of the radiation with which it interacts, recalling that no chemical substance is susceptible to the frequencies outside its characteristic spectrum of frequencies determined by its molecular structure.
	
\subsection{The algebra of frequency matching}
As pointed out in~Section~\ref{WienK}, the distribution function depends on the matching of the two frequency spectra that characterize both the radiation \ul{and} the chemical substance. By adopting the symbols of set theory algebra, let us denote by $\mathbb{R}$ the frequency spectrum of the environmental radiation, and by $\mathbb{L}$ the frequency spectrum of the chemical substance in consideration.

We then define the \textit{susceptibility} $\mathbb{S}$ of the given substance to this particular radiation by the set \textit{intersection}, $\mathbb{S}=\mathbb{R}\cap \mathbb{L}.$ 

When the $\mathbb{S}\ne\emptyset$, we say that the substance is \textit{susceptible} to radiation~$\mathbb{R}$, otherwise, that it is \textit{transparent} to it.

\RMK{The set operations between generating functions are presented in Appendix~\ref{MatrxSet}.}

\subsection{Metastable States}
When the black-body radiation interacts with a chemical substance the latter will be eventually conducted to its fundamental state of \textit{stable} equilibrium. A different radiation source may not cause any change in the thermodynamic system or may conduct it to some \textit{metastable equilibrium state}.

\RMK{The specification of the radiation spectrum is particularly important when the chemical substance can be found in a variety of metastable states, as it happens with clouds in the atmosphere. Meteorological observation revealed that dry air is transparent to most frequencies of solar radiation, though sensitive to the heat emitted from the surface of the earth, that functions as a geographically distributed set of frequency transducers, thus leading to the multiplicity of metastable cloud formations observed.}

\RMK{Note that the steam $pVT$ data, represented in fig.~\ref{FAB1}, does not denounce the metastable states known to exist in water.}

\subsection{The principle of reciprocity}\label{ProcRecip}
According to the \emph{Principle of Conservation of Complex Action} (\pc{CvtvAct}), we can conclude that the {absorption and emission processes} are reciprocal.
	
Besides justifying the Bose character of equations \REF{CA:ops1} and \REF{CA:ops2}, this principle allows developing the analogical reciprocity between the liquid phase of the perfect vapor and radiation, characterized by the \textit{Planck Resonator}, \textsf{i. e.}, the quantum harmonic oscillator.
	
However, due to its abstract character, we are allowed to specify neither the molecular structure of the radicles of the perfect vapor nor the structure of the clusters they engender in the liquid phase. Nevertheless, by assuming the hypotheses advanced by Einstein in his 1917 paper, we can conclude that the processes in which photons and the elements of its phases interact are reciprocal, according to~\pc{xgQuantum}.

\rmk{Part:2}{Considering the mutual independence of the {absorption and emission processes}, equilibrium is settled only when the equation \REF{AA*} is satisfied\footnote{The symbol ``$\approx$'' is here used to denote the fluctuation between the values of the quantities involved.},
\begin{equation}\label{protoEq}
\frac {\avg{\Delta\alpha}} {\avg{\Delta\alpha^*}}\approx 1.		
\end{equation}}

With these considerations, we can take a step towards the statistical equilibrium of the population of particles in the liquid phase of the perfect vapor.

\subsection{Conjectures on the Equilibrium}

As already shown~\CITE{CZM}:

\begin{quotation}\small\it
It is the persistent action of the stochastic processes of {absorption and emission} that lead the quantum system to its equilibrium state.
\end{quotation}

The variations \REF{Ldt} of the actions are assumed to correspond to the three mutually independent stochastic phenomena involved in the {absorption and emission of a photon}, or \textsf{vice versa}, depending on the system in focus, whether radiation itself or the gas of material particles, characterized respectively by the variation of their actions,
\[
\Delta\alpha   = \Delta \alpha_q \qquad\hbox{\rm and}\qquad
\Delta\alpha^* = \Delta \alpha_p,
\]

Hence the effects of the elementary processes of quantum exchange on the state of motion of the gas will be neutralized only statistically, when the \textit{average actions} accumulated during these processes compensate each other{, \textsf{i. e.}, when they attain the following condition of stochastic equilibrium, which, according to the Law of Large Numbers, is attained only after the relaxation time has elapsed:
\begin{equation}\label{AA*}
\avg{\Delta\alpha}_{\rm Relax}=\avg{\Delta\alpha^*}_{\rm Relax}.
\end{equation}

\begin{quotation}\small\it
Since expression \REF{AA*} describes the thermodynamic equilibrium, it will be used in Section~\ref{EqFrmProcs} to evaluate the entropy of the perfect vapor.
\end{quotation}

We can then summarize the results above:

\begin{quotation}\small\it
The connection of Planck's constant $h$ with the imaginary part of the complex representation of the variation of action, gives physical meaning to both the constants, $h$ and the imaginary unit $\imath$, revealing the quantum-relativistic character of the elementary processes of quantum exchange, that correspond to the {absorption and emission of quanta} by the material particles, mathematically described by the non-commutative operators of the creation and annihilation.
\end{quotation}

It is known that the equilibrium formation between the gaseous and the liquid phases in the vapor state of a substance, requires them to be under the action of the processes of absorption and emission of quanta that occur only when this substance is in thermal interaction with a heat reservoir. It is known that to attain its equilibrium state, radiation must undergo the recurrent processes during which photons are created and annihilated.

From the standpoint of the present approach, during the processes of absorption and emission of quanta, the photons are neither annihilated nor created, but merely transferred from the substance to the radiation or \textsf{vice versa}, respectively.
We propose here a tentative description of the gaseous phase of the perfect vapor as a system composed of certain entities which, in their turn, contain a random number $\rj>1$ radicles of the perfect vapor ``\textit{entangled}'' with each other, thus forming a many-body system.

\subsection{Random Variables in the Liquid Phase}\label{clusterXliquid}
An image of the gaseous phase of the perfect vapor can then be depicted as a random number {\bf L} of clusters, where the cluster $\ell$ is a ``molecule'' composed of a random number $\rj$ of radicles of the perfect vapor, characterized by a quantum number $\ell$. The number of radicles in the gaseous phase, in each instant, is therefore a random number ${\bf S_{\bf L}}$, given by the sum,
\begin{equation}\label{SumL}
{\bf S_L}={\mathbf{r}}_1+{\mathbf{r}}_2+\cdots+\rj\cdots+{\mathbf{r}_{\mathbf{L}}}.
\end{equation}
	
\begin{quotation}\small\it 
Equation \REF{SumL} characterizes a compound process. If we denote by 
\[ \beta(s)=\beta_1(s)\cdot\beta_2(s)\cdots\beta_{\mathbf{L}}(s), \]
the comvolution of the generating functions of the random variables $\rj$, then the \textsc{gf} of the random variable, ${\bf S_L}$, is given by $g(s)=g(\beta(s))$~\CITE{Feller}.
\end{quotation}

We then state the following proposition:

\begin{quotation}\small\it
To describe the formalization of the abstract system of bosons, in terns of the random variables {\bf L} and ${\bf r}_\ell$, in agreement with de the description of the current tenets of statistical mechanics, \ul{\sc before} apply it to the perfect vapor.
\end{quotation}

According to the definition \pc{tVap}, this function is an exclusive function of $\theta$. Hence, for $s=\theta$, we have,
\begin{equation}\label{gDEs}
g(s)=g(\beta(\theta)).
\end{equation}

\subsection{The compound generating function of the liquid phase}\label{sinhF}
The generating function $g(s)$ of the random variable ${\bf S}_{\bf L}$, according to the theory of compound distribution processes, is given by the generating function of the liquid phase of the perfect vapor, namely, the extended Boltzmann entropy, \REF{lnAct:E/A1}.

Due to the equation \REF{UVm} that defines the entropy of the perfect vapor for $\varsigma=h$, we conclude that the quantity $\vartheta$, defined in \REF{FGC:vartheta}, must be expressed in terms of a function $\vartheta(\theta)$ of $\theta$. Hence, the relation \REF{lnAct:E/A1} can be written in the form,
\[
\arctan\left(\vartheta_j\right)
%=\frac\imath 2{\rm Ln}\frac
%{1-\imath\vartheta}
%{1+\imath\vartheta}
%=\frac\imath 2{\rm Ln}\frac
%{1-\imath\Phi(\theta)}
%{1+\imath\Phi(\theta)},
=\arctan\left(\vartheta(\theta)\right).
\]

\subsection{Migration of radicles between phases}\label{MR:rec}
In the finite time during which the perfect vapor exchanges a quantum with radiation, one of its molecules, under the action of the laws of the creation-annihilation operators can be either exchanged between the phases $\mathfrak{G}$ and $\mathfrak{L}$ or remain in the liquid phase. Such processes are recurrent, leading the perfect vapor to equilibrium that occurs when absorptions and emissions compensate each other.

The interpretation of the creation-annihilation operators cannot, therefore, be taken literally as the agents of a process during which a particle is created from nothing, or annihilated to nothing, but the agents that cause the \textit{exchange of a particle between the $\mathfrak{G}$ and the $\mathfrak{L}$ phase{s}} (See \pc{DestroiDOFs} and \pc{OnBB:0}).

During an elementary exchange of heat between matter and radiation, the photon that is absorbed by the gas was emitted by the heat reservoir, and the photon emitted by the gas is absorbed by that reservoir. We can then conclude that the occupation of photons in quantum states of radiation correspond \textit{reciprocally} to the occupation of particles of the gas in the liquid phase, the form of the partition function for particles in the cluster of quantum number $\ell$ is the \textit{same} for photons in the quantum state $\ell$, 
\begin{equation}\label{cschM}
\beta(s)\Rightarrow\frac 12\ {\rm csch}\left(\frac\theta 2\right),
\end{equation}
however with a different argument $s$ of the function $\beta(s)$ in equation \REF{csch}, corresponding to particles with mass. We therefore put $s=\theta$, obtaining, as expected from definition \pc{tVap}. Hence, the expression of the {\sc gf} of the random variable ${\bf S_L}$ is given by,
\[
g\left(\frac 12\ {\rm csch}\left(\frac\theta 2\right)\right).
\]

In the particular case in which the perfect vapor is under the influence of black-body radiation, the function $\vartheta(\theta)$ is given by \REF{cschM}, which gives the generating function of the thermodynamic properties of the perfect vapor's liquid phase, that, in its turn, describes a compound process,
	
\begin{equation}\label{E:arctanTheta}
\frac{S_{\mathfrak{L}}}k
%=\arctan\ \frac 1{2\sinh\left(\frac \theta 2\right)}
%=\ {\rm arccot}\ \left[2\sinh\left(\frac \theta 2\right)\right]
=\left\{\left(j+\frac 12\right)\pi-\arctan\ \left[2\sinh\left(\frac \theta 2\right)\right]\right\},
\end{equation}
where the integers $j=1,2,\cdots$ might represent the contribution of some yet unexplained phenomenon.
	
\subsection{The equation of state of the perfect vapor}\label{EOSBg}
From the entropy \REF{E:arctanTheta} it was possible to derive the function,
\begin{equation}\label{fthetaBG}
f(\theta) = \frac\theta 2\sech\left(\frac\theta 2\right),
\end{equation}
which appears in the equation of state \REF{pV=RT}.

As shown in Appendix~\ref{Steam}, after removing the influence of the supposedly contribution of the \textit{nucleation processes} that give rise to the $\mathfrak{L}$ (liquid)-phase from steam $pVT$ data, their adherence to a single curve in the $\theta\times\zeta$ plane (fig. \ref{FAB2}) is noteworthy. It is then justified to confront the theoretical expression \REF{fthetaBG} with these data.

Since the theoretical function $f(\theta)$ is an odd function\footnote{This finding might justify the notion of quasi-particles in dealing with certain many-body problems.}, it can explain the remarkable half-turn symmetry observed in the vapor region of steam. Besides, it reveals that the observed adherence to a single curve, depends on the translation of the origin of \REF{fthetaBG} to the center $[\theta_m,\zeta_m]$ of symmetry in the experimental curve, localized in the mid-point of the segment $AB$. The non-negligible dispersion of points around the point $B=\left[\theta^*,\zeta^*\right]$, and the low accuracy method used for its determination impedes the accurate location of this point in the graphics.

With these empirical evidences, the entropy \REF{E:arctanTheta} of the perfect vapor, can be rewritten in the form,
\[
\frac{S_{\mathfrak{L}}}k
=\left\{\left(j+\frac 12\right)\pi-\arctan\ \left[2\sinh\left(\frac{\theta-\theta_m} 2\right)\right]\right\},
\]
where $j$ is a constant whose presence, although mathematically justified, could not be explained by me.

\appendix
%A
\section{Thermodynamic Properties of Steam}\label{Steam}

This Appendix confronts the $pVT$ data of steam against the equation of state of the perfect vapor, and its departure from the ideal gas law. The remarkable symmetry observed in the graphics here presented justifies further inquire on this subject.

While the indeterminacy of the function $f(\theta)$ endows the perfect vapor with generality, its universality cannot be claimed until an extensive confrontation with experiment is fulfilled, an endeavor that is beyond the scope of this paper.

Since $z$ and $\theta$ are both dimensionless quantities, the plot of the $pVT$ data of any substance in the $\theta\times z$ plane is meaningful for both a thermodynamic and a quantum reading. The closer these data are to a single curve%
\footnote{The quotient obtained by dividing the area occupied by the $pVT$ data, by the total area determined by the selected intervals of $z$ and $\theta$, gives a rough measure of how close the substance is to the perfect vapor.}, the better the substance can be represented by equation \REF{pV=RT}, where $f(\theta)$ represents the departure of the perfect vapor from the Clapeyron equation. 

\subsection{The departure of steam from the ideal gas}
To form a rough idea of the functional form of $f(\theta)$, the graphic of $\theta\times z$ of the steam $pVT$ data\footnote{Figures \ref{FAB1} and \ref{FAB2} were introduced here for illustration purposes only. The data there exhibited were obtained from an old steam table I had at hand~\CITE{Haywood}, when I programmed and used the algorithms to obtain the value of the parameter $p_\ell$, required to produce fig. \ref{FAB2}. Since its determination was based on a low accuracy visual trial and error procedure, higher accuracy in the approximation \REF{Nucl:1} would be of little value.}
is exhibited in fig. \ref{FAB1}.
\begin{figure}[h]
\begin{center}
\begin{tabular}{cc}
\includegraphics[width=15cm]{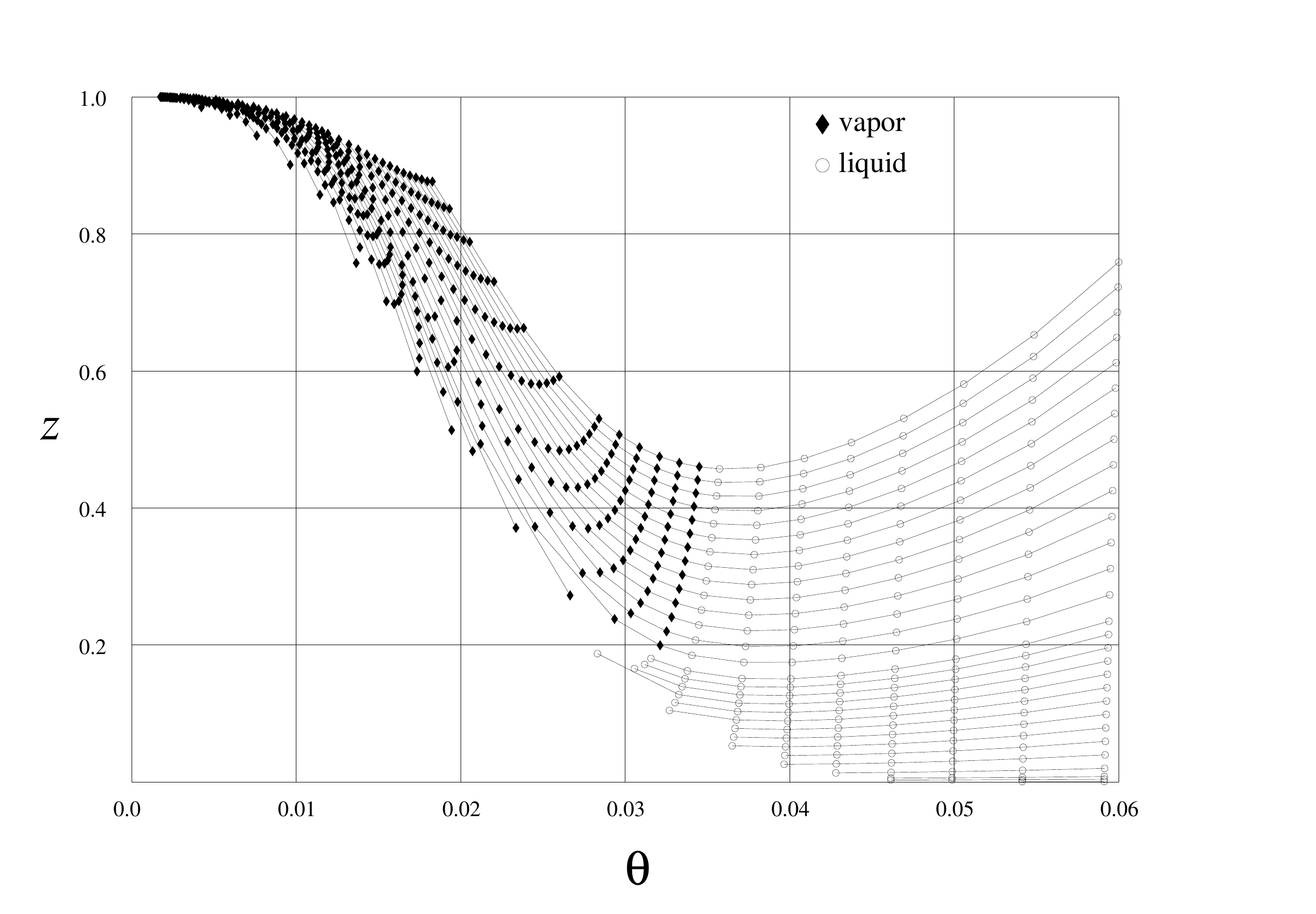}
\end{tabular}
\caption{\label{FAB1}
$pVT$ data for steam in the $\theta\times z$ plane.}
\end{center}
\end{figure}

While the isobaric curves of steam do not meet in a single curve, they form a family of isomorphic shapes, regularly displaced in the vertical direction, I was suggested to believe that the departure of the steam $pVT$ data from the ideal gas law is due to the composition of at least two phenomena: on one hand, the displacements explained by the perfect vapor, and on the other hand, the contribution of an independent phenomenon that causes the departure of the isobaric curves from the perfect vapor. 

Considerations about the causes of the formation of the entropy of steam led me to assume that steam is described by the superposition of the Clapeyron-Clausius entropy variation during a phase transition to the perfect vapor behavior, supposedly described by the clustering of water molecules\footnote{This hypothesis is justified by its consequences.}. I therefore propose the following conjecture:

\RMK{The departure of the isobaric curves from the perfect vapor are due to nucleation processes, and that nucleation and clustering are independent processes.}

\subsection{The departure of steam from the perfect vapor}\label{K:1}
To form a clearer idea of the functional form of $f(\theta)$, it would be necessary to eliminate the influence of the unknown cause of the departure of the steam from the perfect vapor.

It is here conjectured that such departure is due to the nucleation phenomenon, by assuming that the vapor pressure $p$ of a substance is given by the equation,
\[
\frac{p}{p_\ell}=e^{-\frac{\Delta H}{RT}},
\]
where $\Delta H$ is the variation of the enthalpy of the system during a vaporization/condensation process, and $p_\ell$ is a constant pressure, characteristic of the liquid state.

For small values of $\Delta H/{RT}$, we can write $\exp(-\Delta H/RT)\approx 1-\Delta H/RT$ to express the energy equation,
\[
\frac p{p_\ell}{RT} + \Delta H\approx RT=\frac p{p_\ell}{RT} + \Delta U+p\Delta V\approx RT.
\]
Since in a change of state $p\Delta V=V-V_\ell$ where $V$ is the volume of gas and $V_\ell$ of the liquid, we can write $p\Delta V\approx pV$, thus obtaining,

\begin{equation}\label{Nucl:1}
pV+\frac p{p_\ell}{RT} + \Delta U\approx RT,
\end{equation}

\begin{figure}[h]
\begin{center}
\begin{tabular}{cc}
\includegraphics[width=15cm]{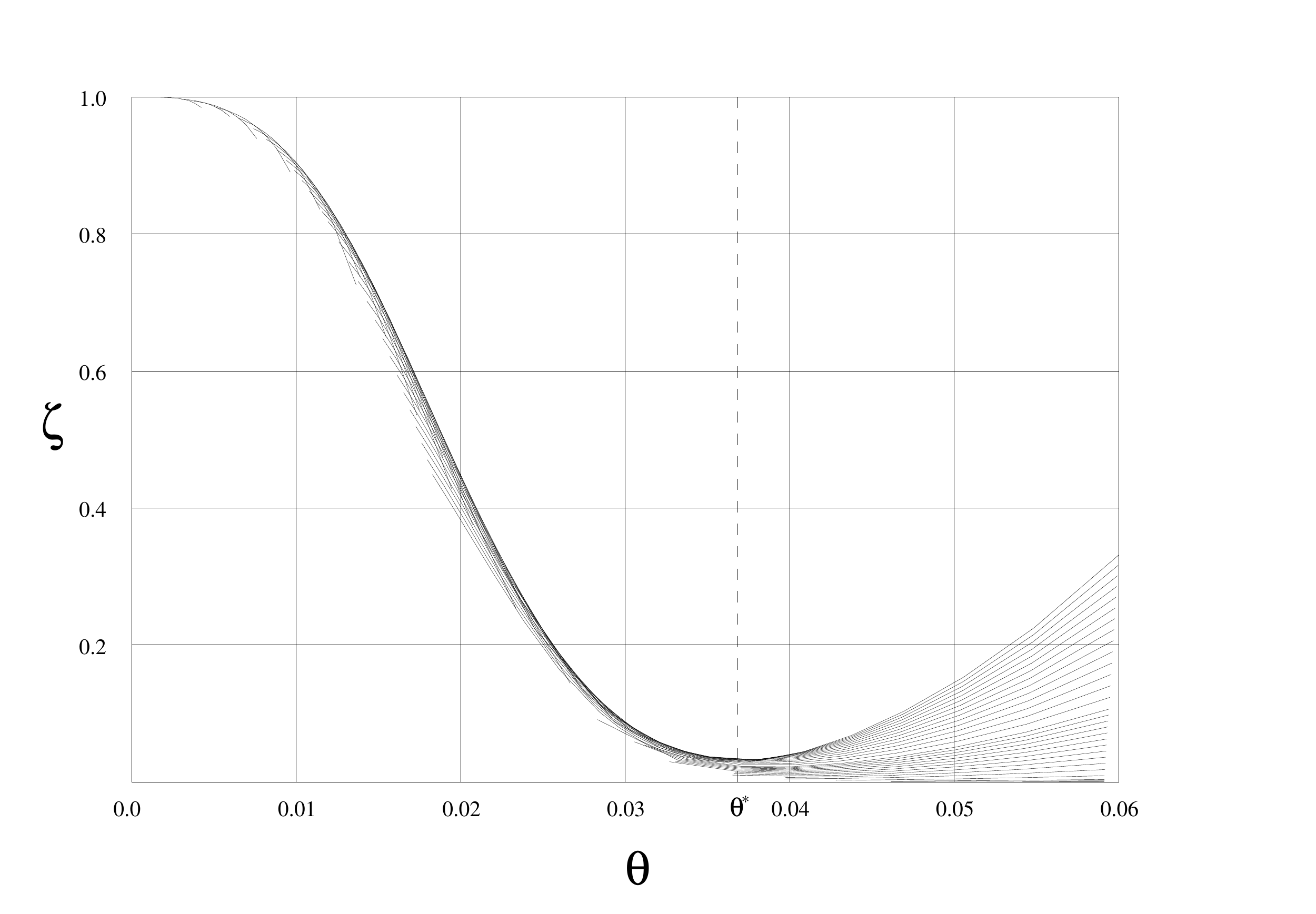}
\end{tabular}
\caption{\label{FAB2}
$pVT$ data for steam in the $\theta\times\zeta$ plane.}
\end{center}
\end{figure}

{Comparing the quantities $RT-pV$ obtained from equations \REF{pV=RT} and \REF{Nucl:1}, we conclude that they represent different departures from the Clapeyron equation, due to distinct phenomena. By assuming that the behavior of the steam is the result of their combined action, we obtain its equation of state in the following form,
\begin{equation}\label{Nucl:2}
z+f(\theta)+\frac p{p_\ell}\approx 1.
\end{equation}}

Fig. \ref{FAB2} represents the steam data in the $\theta\times\zeta$ plane%
\footnote{The value of $p_l\approx 2341$ bar in equation \REF{Nucl:2} was obtained by a low-accuracy visual trial and error method, that made the consideration $L/RT$ superfluous for a reliable estimation of $p_\ell$.},
where,
\begin{equation}\label{pp0}
\zeta=z-p/{p_\ell}.
\end{equation}
Without the influence of nucleation on the properties of steam, the pressure dispersion becomes largely attenuated, so that \emph{all} isobaric curves inside the vapor region become confined in a remarkably narrow belt upper-bounded by a single, sharp limiting curve.

The determination of the values of $\theta^*$, $z^*$, and $p_\ell$, that requires further knowledge about steam, were merely guessed for the construction of fig. \ref{FAB2}. Hence, this graphic cannot be considered as conclusive, but as merely suggestive of the potential of the present approach.

\subsection{Remarkable symmetries in steam}\label{Simetrias}

Some aspects of the upper-bound curve in this graphic are noteworthy. The experimental data closely adheres to a single curve in the whole vapor region. As opposed to the small deviations observed under it, this curve is extraordinarily sharp.

Let us denote the maximum and minimum of the curve, respectively, by $A$ and $B$, whose coordinates in the $\theta\times\zeta$ plane are $A=[0,1]$, and $B=\left[\theta^*,\zeta^*\right]$. These two points delimit the vapor region\footnote{\label{Hyperbolic}It can be seen that the value of $\theta$ at $B$ separates the vapor from the liquid phase of water.}.
It can be clearly seen that around the mid-point of the segment $AB$ the curve exhibits a remarkable half-turn symmetry.
It can be seen that, the greater the value of $\theta$ in the vapor region, the more degenerate is the gas and the closer it is to its liquid state. It is interesting to confront these symmetries against those that arise in the quasi-particles representation of collective phenomena.

%B
\section{The Kinetics of Perfect Gases}\label{qKinetics}

This Section presents a summary of the results obtained by the author~\CITE{CZM}\ in the treatment of a gas whose state is given by the population of particles in its enumerable set of quantum states. These numbers are treated as random variables, whose \textsc{pdf}s are the solutions of the general difference-differential equation of Markovian birth-and-death stochastic processes, whose \textit{laws of change} derive from the reciprocal theorem of many-body theory, where from the indistinguishability and Pauli's exclusion principles have been hitherto derived.

These solutions describe the transient evolution of the gas towards the thermodynamic equilibrium, that reproduces the results already obtained for the equilibrium state of the gases of Bose and Fermi particles, that can be extended to a gas of Boltzmann particles.

Besides predicting the time asymmetry previewed by the second law, it reveals that the indistinguishability of identical particles and Pauli's exclusion rule are not conditions of necessity, but of equilibrium. Hence, the particles of a gas removed from equilibrium are not said to be indistinguishable, but rather to \textit{become statistically indistinguishable} in the state of equilibrium.

\subsection*{Population of quantum states as dynamic variables}
It is known that for high temperatures and low densities the gases of Bose particles (the black-body radiation) behave as a perfect gas, characteristic of the gaseous phase of the perfect vapor. For lower temperatures and higher densities the quantum phenomena become increasingly relevant. The purpose of this Section is to study the influence of the liquid phase on the formation of thermodynamic equilibrium. We will then focus on the time evolution of the random occupation numbers,
\[
{\mathfrak{R}}(t)=\{\rjj{0}(t), \rjj{1}(t), \rjj{2}(t), \cdots, \rj(t), \cdots\}
\]
of the corresponding enumerable set of wave functions,
\[
\Psi=\left\{\psi_0,\psi_1,\psi_2,\cdots,\psi_\ell,\ldots\right\}
\]
that characterize the clusters $\ell=0,1,\cdots$ that compose the liquid phase of the gases of Boltzmann, Bose, and Fermi particles, as described by the Second Quantization formalism.

{Recall that, according to quantum theory that, amended by this approach, the average values, 
\[
\bar\rjj{\ell}=\lim_{t\rightarrow\infty}\rjj{\ell}(t),\ \ell=0,1,2,\cdots
\]
are given by the eigen-values of the dynamic variables under the persistent action of the {ladder operators}}. The time variable $t$ is here defined according to the Markovian stochastic processes~\CITE{Feller}. 
	
\rmk{avg}{\rm The current interpretation this axiom of quantum theory evokes a whimsical comment of Szent Gy\"orgyi when he joined the Institute for Advanced Study in Princeton about the behavior of more than two electrons\footnote{Cited by von Bertalanffy~\CITE{Bertalanffy}.}:}

\begin{quote}\small\it``{I did this in the hope that by rubbing elbows with those great atomic physicists and mathematicians I would learn something about living matters. But as soon as I revealed that in any living system there are more than two electrons, the physicists would not speak to me. With all their computers they could not say what the third electron might do. So that little electron knows something that the wise men of Princeton don't, and this can only be something very simple.}'' \end{quote}

While talented humans can devise algorithms to calculate \emph{with all their computers} the eigenvalues and eigenvectors specified in Axiom~\ref{frAxiomD}, photons can not. Since the operators {\textsf{a}}, and {\textsf{a}}$^\dagger$ act independently in nature, we are not allowed to conclude that photons are endowed with the algorithmic faculties that computers have to calculate that eigen-stuff.

What is here required is to prove that the eigenstate (the equilibrium state) of a dynamical variable is attained by the recurrent action of the conditional probabilities of the elementary quantum state transitions, described by the {ladder operators}, that are mutually time independent. This process is neither deterministic, in the sense of regular convergence, but instead, stochastic\footnote{Or, if preferred, a Monte-Carlo algorithm.}. This approach supersedes the \textit{Principle of Molecular Chaos} adopted in statistical mechanics:

\rmk{Eq:Mb}{To derive the probability distribution functions of the occupancy numbers of the quantum states that characterize the liquid phase of the perfect vapor, from the \textit{laws of change} described by the \textit{conditional probabilities} laws of creation and annihilation.}

Hence, it is reasonable to assume that the evolution of these transient transitions must be described as conceived in probability theory,
\begin{quote}\small\it
A ``{conceptually (\ldots) analogue of the processes of classical mechanics, where the future development is completely determined by the present state and is independent of the way in which the present state has developed}.''\footnote{p. 420,~\CITE{Feller}}. 
\end{quote}
Such analogue is the Markovian birth-and-death processes, whose \emph{laws of change} --- that correspond to the \emph{laws of force} of Newtonian mechanics --- are given by the creation and annihilation operators.

The difference between the current tenets of statistical mechanics and the approach here introduced is subtle. While in the former, time is explicit in Schrödinger's time only equation, in the latter it is intrinsic to the Markov equation. The main consequences of this difference are that (a) the latter previews time asymmetry, while the former does not, and (b) in the latter, indistinguishability and Pauli's exclusion principles are interpreted as conditions of equilibrium imparted to the system of particles by the way they are shuffled, while in the former these principles assume that the behavior of these particles is due to their ontological nature.

We then have to face the following dilemma:

\tpk{QH:6}{Whether we assign to the creation--annihilation operators a secondary and passive role in quantum theory, as a mere mathematical curiosity designed to help us in our resignation to accept the bizarre behavior of these particles, or we recognize in them a \textit{primary} \textit{active role} in the shuffling process, as the \textit{elementary chaos} that cause that behavior. In the words of the philosophers of Ancient Greece, they are to be seen as the \emph{unchanging principles of change}, that gives physical meaning to the \emph{clinamen} (swerve) of Atomistic doctrine~\CITE{Lucretius}.}

\subsection{On being and becoming} \label{dilema}
\begin{flushright}\small\it
Far away from saying that the object precedes the point of view,\\
we would say that it is the point of view that creates the object.
\vskip10pt
{\em F. Saussure, {\sl Cours de Linguistique G\'en\'erale}} (1916).
\end{flushright}
Being unconcerned with the accidental circumstances of an actual experiment, the sample space approach to random phenomena considers neither the {\em initial configuration} of its moving parts (balls, dice, coins, cards or particles) nor the changes they undergo {\em during} the trial. These considerations suggest that the replacement of the kinetic theoretical methods by those of statistics of the sample space, resulted in what might be a misleading epistemological shift that can be described {by Saussure's aphorism reproduced in the heading of this Section}: by focusing on the outcomes of the irreversible processes instead of on the processes themselves, one might be subtly persuaded to interpret \textit{indistinguishability} and \textit{exclusivity} as essential characters of the particles, instead of a mere effect of shuffling on their outcomes.

Although Saussure's approach to linguistics is a conscious epistemological decision, the choice of the sample space approach by the statistical mechanics is a convenient though misleading simplification.

In fact, any attempt to derive the properties of a phenomenon that depends on the process of shuffling from the statistical analysis of its outcomes might be illusory, as evidenced by the following abstract example.

\subsection*{Sample space mirages}
Let us consider the random placement of $r$ balls in $n$ cells\footnote{This is an imaginary shuffling process. It should not be taken as a model for the molecular chaos in a gas.}.
\begin{quote}\small
Suppose that the balls are made of ice, each one with the same mass $m$ and frozen into a different mold, so that they can be numbered, thereby being \textit{distinguishable} by any observer. Let us suppose also that $n=26$, and that each cell is labeled with one letter of the alphabet, so that any point in the sample space of such experiment can be uniquely identified with a word of length $r$: the first letter denotes the cell in which ball 1 was placed, the second, the cell containing ball 2, and so on%
\footnote{Think of the sequential placement of balls in the cells as the sequence of the keys (cells) you stroke (put the ball) in a keyboard to produce a word.}.

Let us also suppose, as it happens with the particles in a gas, that the shuffling is nonobservable and performed below the fusion temperature of the water. Hence, any observer can read, in the outcome of each trial, a word of $r$ letters. He then concludes that the ice balls are distinguishable.

Let us suppose now another shuffling performed at a higher temperature and lengthy enough to allow the balls to melt during the relaxation process. As time goes on, the shuffling process will, as it were, act no longer on solid balls, but instead on melted balls. The balls, originally distinguishable, \emph{become} indistinguishable. In fact, looking at the outcome of the shuffling, the observer will be unable to read, in the configuration he sees, a word of length $r${, \textsf{i. e.}}, he cannot identify the balls in the cells. Instead, the only information he can draw from his observation --- by weighting each cell to know the mass of water it contains --- is given by the occupancy numbers, \textsf{i. e.}, the number of balls he finds in each cell. Misled by the ``external aspects'' of the experiments (their \emph{phenotypes}), instead of recognizing that the balls {\em became} indistinguishable during the shuffling process, he concludes incorrectly that, since they {\em are} indistinguishable at the moment of the observation, they {\em have always been} so.

A privileged observer of this game (knowing its \emph{genotypes}) knows, as opposed, that he cannot blame the balls for their indistinguishability, for it is the \textit{shuffling process} itself that imparts to the balls such bizarre behavior. Instead, he knows that shuffling engenders a sort of ``\emph{indistinguishabilization}'' process, so that the ice balls, originally distinguishable, \emph{become} indistinguishable.
\end{quote}

The paradox raised in confronting Einstein's (1917) ``\textit{kinetic}'' against Bose's ``\textit{combinatorial}'' derivation of Planck's radiation law disappears when examined in terms of the elementary processes that, as in the example above, lead a system of Bose particles to its final state of equilibrium.

Hence, indistinguishability need not be understood as an ontological permanent characteristic of a system of identical particles. It can, instead, be seen as the consequence of the way its particles are shuffled, \textsf{i. e.}, as the way they interact with radiation. In fact, when the gas is removed from equilibrium by a finite thermodynamic transformation\footnote{A thermodynamic transformation is understood as a large scale departure from equilibrium that removes the system far away from the statistical fluctuation domain.}, Einstein's equilibrium equation no longer holds, \textsf{i. e.}, the symmetry of its wave function is necessarily broken\footnote{Einstein's equilibrium equation can be obtained as the asymptotic form of the kinetic equation of gases~~\CITE{CZM}.}.

As a corollary of Einstein's \textit{kinetic} derivation of Planck's radiation formula, we can state:

\tpc{notnec}{Indistinguishability and equilibrium}{Indistinguishability is not a condition of necessity, but of equilibrium.}

The same holds for Pauli's exclusion principle.

\subsection{Transient processes in the perfect vapor}\label{GasDynamics}
While the behavior of the gaseous phase of the perfect vapor has been satisfactorily described in kinetic theory\footnote{\label{fnTrans}While some authors assume that the translation of a molecule in a fixed volume is quantized, its supposedly discrete energy levels have not yet been measured by spectroscopy. In the domain \REF{ineq} of the phase space, the particles do not interact with radiation and therefore, cannot be supposed to exhibit quantum behavior.}, the description of the liquid phase requires a quantum approach. Hence, to give a kinetic approach to \textit{condensation phenomena} we shall extend the theoretical definition of a \textit{many-body quantum system} to fit the \textsc{epqe} that conduct the particles of the perfect vapor to its thermodynamic equilibrium.

Recall that a many-body quantum system is composed by a number $n>1$ of particles \textit{entangled} with each other as the result of a chemical reaction that destroys their degrees of freedom and merges them by their differential Hamiltonian operator into a single wave function. According to the current tenets, every wave function is subject to the symmetric and antisymmetric constraints, known as the indistinguishability and Pauli exclusion principles. Such definition imposes indistinguishability and exclusivity at the outset.

Nevertheless, quantum axioms allow predicting the existence of state transition phenomena, assigning conditional probabilities to their occurrence. Since the staqte transition probabilities are assigned to the occupancy numbers, the values of such numbers cannot be prescribed; they must, instead, be treated as \textit{random variables}, thereby allowed to acquire arbitrary values not necessarily equal to the value $\bar\rj$, imparted by the symmetric or anti-symmetric constraints.

The way out this contradiction is merely a matter of choice between the two interpretations of the indistinguishability and Pauli exclusion principles expressed in \pc{QH:6}, whether we assume with the current tenets that these rules reveal the ontological character of every many-body system of identical particles, or that they merely describe the condition of \textit{thermodynamic equilibrium}.

In the following it will be shown that, by assuming that the {ladder operators} describe the agents that cause the changes the occupancy numbers of quantum states undergo, it is possible to obtain the time evolution of the quantum states that starts from an arbitrary initial \textsc{pdf} until it attains the \textsc{pdf} characteristic of thermodynamic equilibrium.

The $\rj$'s are random variables whose behavior can be likened to the placement of balls (particles) in cells (quantum states).

The replacement of the ideal gas by the perfect vapor to represent real gases, that recognizes the existence of a random number of components, as opposed to the misleading mirage discussed in Section~\ref{dilema}, leads to a far-reaching approach to its ``laws of motion'', that are described, not in terms of the \textit{trajectories} of its particles, but in terms of the time evolution of the \textsc{pdf} of the \textit{occupancy random numbers} of the quantum states of the many-body system in question.

\subsection{The kinetics of the perfect vapor}\label{gasdynamics}

To represent the motion of the gas it is more realistic to assume that during a short interval of time, not all, but only a small number of particles change their positions (whether in the gaseous phase, or in the liquid phase), most of them remaining in the same phase they were at the beginning of the interval. We are therefore justified to state the \textit{continuity hypothesis}% 
\footnote{Since this hypothesis does not refer to equilibrium, it is superfluous to statistical mechanics .}:

\tpc{continuity}{Continuity}{The smaller the time interval considered, the smaller the number of particles changing their states.}

It is known from statistical mechanics that the values of both the expectation and the variance of the random variables $\rj(t)$ in the equilibrium are negligible when compared with the extremely large numbers of particles and quantum states in the perfect vapor. Hence, we are allowed to state the \textit{independence hypothesis}%
\footnote{\label{SM}%
Statement \pc{independence} is here treated as a \textit{proposition}, in spite of being tacitly stated as true in the method of \textit{most probable distribution} of statistical mechanics, where from its preamble derives. It is not possible to decide, at this point, whether it describes the general behavior of molecules, or merely represents a complement to the definition of ideal gases, as expressed in \REF{eq:rates}.}:

\tpc{independence}{Independence}{The removal of any cluster (together with the radicles it contains) from the perfect vapor, will not modify the flow processes that take place in the remaining clusters. In other words, the flow of radicles in a given cluster is independent of the flow that occurs in any other cluster%
\footnote{In fact, this flow depends exclusively on the population of the state and on the properties of radiation.}.}

With these assumptions, the investigation of the laws of motion of the perfect vapor is therefore reduced to find the laws that rule the arrival and departure rate, to and from, a single cluster.

To describe this flow let us denote by $A_\ell(\Delta t)$ and $D_\ell(\Delta t)$, respectively, the number of particles arriving to, and departing from, the state $\psi_\ell$ during the time interval $\Delta t$. Let us consider their time power series expansion,
\begin{eqnarray}
A_\ell(\Delta t) &=& A_\ell(0) + \dot A_\ell\Delta t + \ldots \nonumber\\
D_\ell(\Delta t) &=& D_\ell(0) + \dot D_\ell\Delta t + \ldots \nonumber
\end{eqnarray}
where $\dot A_\ell(t)$ and $\dot D_\ell(t)$ are, respectively, the arrivals and departure rates of state $\psi_\ell$. According to hypothesis \pc{continuity}, we have $A_\ell(0) = D_\ell(0) = 0$. Therefore the first-order equilibrium condition of the flow is given by,
\begin{equation} \label{eq:equilibrium}
\dot A_\ell(t)=\dot D_\ell(t).
\end{equation}

It can be easily verified that both the hypotheses \pc{continuity} and \pc{independence} are subsumed in the following axiom that determines the \textit{law of motion} of the {\sc pdf} of the occupancy number of a given state:

\rmk{MarkovianFlow}{The flow of particles in a quantum state is a Markovian birth-and-death process.}

Recall that the master equation for ${\bf P}_{\rj}(t)$ of a Markovian birth-and-death process is given by the difference-differential equation~{\CITE{Feller}},
\begin{equation} \label{MarkovI}
\frac {\partial {\bf P}_{\rj}(t)}{\partial t} =
\lambda_{\rj-1}\,{\bf P}_{\rj-1}(t)
-\left(\lambda_{\rj}+\mu_{\rj}\right){\bf P}_{\rj}(t)
+ \mu_{\rj+1}\,{\bf P}_{\rj+1}(t),
\end{equation}
where $\lambda_{\rj}$ and $\mu_{\rj}$ are, respectively, the arrival and departure rates that describe the laws that rule the shuffling processes. To complete the description of the motion of the gas, the knowledge of the laws that determine these rates are required.

\rmk{time}
{In this approach, Planck's constant acknowledged as an imaginary entity, $\imath\hbar$, thus assigning a correspondingly different meaning to \emph{time}, derived from the Minkowski space-time description of special relativity. The time partial operator $\delp{}t$ is then detached from Schrödinger's equation, and superseded by the Markov difference-differential equation, treated independently from Schrödinger's amplitude equation, that, according to Section~\ref{TheEquil}, is understood as a descriptor of the thermodynamic equilibrium condition.}

The values of the arrival and departure rates for the gases of Fermi, Bose, and Newtonian particles are given by~\CITE{CZM},
\begin{equation}\label{eq:rates}
\lambda_{\rj }=\lambda\left(1+\beta\rj \right)\quad\hbox{and}\quad
\mu_{\rj}=\mu\rj .
\end{equation}
where $\lambda=\xi e^{-\epsilon_\ell/kT}, \mu=\xi e^{-\eta/kT},$ and $\xi$ is an unknown frequency rate. From the creation-annihilation operators of second quantization, we have $\beta=1$ for Bose, $\beta=-1$ for Fermi particles. The value $\beta=0$ results in the Maxwell-Boltzmann {\sc pdf}.

Let us denote by $\Pi_\ell(z,t) = \sum_{\rj} {\bf P}_{\rj}(t)z^{\rj},$ the generating function ({\sc gf}) of ${\bf P}_{\rj}(t)$. After substituting the \textit{laws of change}~\REF{eq:rates} in~\REF{MarkovI}, we arrive at,
\begin{equation}\label{eq:Master}
\lambda(z-1)\Pi_\ell = \frac{\partial\Pi_\ell}{\partial t}
-(z-1)(\beta z-\mu)\frac{\partial\Pi_\ell}{\partial z},
\end{equation}
whose solutions can be written in the general form,
\[
\Pi_\ell(z,t) = \varphi_\ell(z)\cdot \Phi_\ell(z,t),
\]
where $\varphi_\ell(z)$ is the {\sc gf} of the average occupancy numbers $\bar\rj$ of the state $\ell$ in the equilibrium, and $\Phi_\ell(z,t),$ is the {\sc gf} of the transient population $\mathbf{y}_\ell(t)$ in that state, at time $t$. According to the convolution theorem, the population of state $\ell$ at time $t$ is $\rj(t) = \bar\rj + \mathbf{y}_\ell(t).$

\subsection{Probabilistic dynamics of particles}\label{trajectories}
A parallel between the equation \REF{MarkovI} and Newton's second law can help elucidate the nature of the former. The Newtonian differential equation $\dot p=f$ remains indeterminate until a \textit{law of force} $f(r)$ that is independent of the postulates and axioms of mechanics, is provided. The mathematically indeterminate role played by the law of force in Newtonian mechanics, is played by the correspondingly indeterminate \textit{laws of change} (the arrivals and departures rates) in quantum mechanics. Both lead to the laws of evolution: the former giving the trajectory of the particle, the latter, the time evolution of the {\sc pdf} of the occupancy numbers of the states of the quantum state. Both require the specification of initial conditions.

\subsection{Further Comments on Indistinguishability}\label{fCmts}

Equation \REF{MarkovI} is abstract enough to be in the foundations of a \textit{General Theory of Change}, that can be applied to non-thermodynamic phenomena. Equation \REF{eq:Master} is a confirmation that indistinguishability can be legitimately interpreted as a statistical phenomenon that can be represented by the random placement of balls of whatever nature (billiard balls, for instance) in cells, provided the \textit{shuffling} is performed according to appropriate \textit{laws of change} that ``indistinguishabilize'' the balls. Indistinguishabilization can, therefore, be seen as a random process whose effect is to induce a particular partition on a set. As a mere consequence of the \textit{shuffling} in a birth and death process, indistinguishability is therefore expected to be found not only in quantum phenomena but also in the population dynamics of systems of elements whose nature might be quite unlike the physical elementary particles dealt with in quantum physics and statistical mechanics.

It is worthwhile to note that most of the requirements that, in the view of \textit{Huang}, a ``satisfactory'' derivation of statistical mechanics  should fulfill~\CITE{Huang}, are satisfied by the present approach to the gas dynamics described in Section \ref{gasdynamics}:
\begin{quote}
\begin{enumerate}\sl
\item a not-\textit{ad hoc} assumption of the molecular chaos by reducing it to a first-principles representation, given in terms of \textit{natural laws of change};
\item a detailed description, at least for the case of ideal gases, of the approach to equilibrium;
\item a master equation~\REF{MarkovI} expressed, not in terms of the wave function, but of the {\sc pdf}s of occupancy numbers of the states of the ideal gases.
\end{enumerate}
\end{quote}

%C
\section{Matrix and Set Operations with Generating Functions}\label{MatrxSet}

This Appendix presents an extension of operational methods with generating functions to matrix and set theory operations.

\paragraph{Generating Function.}The generating function GF of a matrix ${\bf A}=a_{ij}$
is given by the polynomial,
$$
A(x,y)=\sum_i\sum_j a_{ij}x^iy^j.
$$
where $x$ and $y$ are dummy variables.

\paragraph{Theorem.}Let ${\bf A}=\{a_{ij}\}$ and ${\bf B}=\{b_{ij}\}$ be two matrices.

The GF of the matrix ${\bf C}={\bf A} \times {\bf B}$ can be
obtained by the evaluation of the contour integral
\begin{equation}\label{prod}
C(x,y)= \frac{1}{2\pi \imath}\oint A(x,\zeta)
B\left({\frac{1}{\zeta}},y\right)\frac{d\zeta}{\zeta}
\end{equation}

\paragraph{Proof.} Let $A_j(x)=\sum_i a_{ij}x^i$ and %
$B_i(y)=\sum_j b_{ij}y^j$.
Substituting these terms in (\ref{prod}), we obtain
$$
\begin{array}{ll}
C(x,y)&=\frac{1}{2\pi \imath}\oint \left\{\sum_k A_k(x)\zeta^k\cdot
        \sum_m B_m(y)\frac{1}{\zeta^m}\right\}\frac{d\zeta}{\zeta}   \\
      &=\sum_k \sum_m A_k(x)B_m(y)
\left\{\frac{1}{2\pi \imath}\oint\frac{d\zeta}{\zeta^{k-m+1}}\right\}
\end{array}
$$

Since
$$
\oint\frac{d\zeta}{\zeta^{k-m+1}}=\delta_{mk}=\left\{
\begin{array}{l}
0 \quad \hbox{if} \quad m\neq n\\
1 \quad \hbox{if} \quad m = n
\end{array}
\right.
$$
then
$$
\begin{array}{ll}
C(x,y)&=\sum_k A_k(x) B_k(y)\\
      &= \sum_k\left\{\sum_i a_{ik}x^i\cdot \sum_j b_{kj}y^j  \right\} \\
      &= \sum_i \sum_j \left\{\sum_k a_{ik}b_{kj}x^iy^j\right\}
         \quad\hbox{\bf q.e.d.}
\end{array}
$$

\subsection*{Matrix Theory}

\paragraph{Identity.}
Comparing (\ref{prod}) with the Cauchy integral
$$
A(x,y)= \frac{1}{2\pi \imath}\oint A(x,\zeta)\frac{d\zeta}{\zeta-y}
$$
we can readily recognize the function
$$
I(x,y)= \frac{1}{1-xy}
$$
as the GF of the identity matrix.

\paragraph{Inverse.} By stating $C(x,y)=(1-xy)^{-1}$ in (\ref{prod}),
we can obtain the GF of the inverse of a matrix {\bf A} as the solution
$B(x,y)$ of the integral equation~(\ref{prod}).

\paragraph{Diagonal.} The GF $A(x,y)$ of a diagonal matrix is a matrix of the
form $A(xy)=A(z)$.

This property extends the application of integral~(\ref{prod}) to operations in
set theory.

\subsection*{Set Theory}

A subset {\bf C} of the non negative integers may be defined by the generating
function
$$
C(z)=\sum c_k z^k, \quad \hbox{where}\quad c_k=\left\{
\begin{array}{l}
1\quad \hbox{if}\quad c_k\in {\bf C} \\
0\quad \hbox{otherwise}
\end{array}
\right.
$$

\paragraph{Universe.} The GF for all non negative integers is $\frac{1}{1-z}$.

\paragraph{Complement.} The GF of the complement $\overline{\bf C}$ of a set
{\bf C} is
$$
\overline{C}(z)=\frac{1}{1-z}-C(z).
$$

Let {\bf A} and {\bf B} be two sets defined as {\bf C} above.

\paragraph{Intersection.} The GF $E(z)$ for the intersection %
${\bf E}={\bf A}\cap{\bf B}$ of the sets {\bf A} and {\bf B}, is obtained
by the integral
$$
E(z)=E(xy)=\frac{1}{2\pi \imath}\oint A(x\zeta)\cdot
B\left(\frac{y}{\zeta}\right)\frac{d\zeta}{\zeta}.
$$

\paragraph{Union.} The GF $U(z)$ for the union ${\bf U}={\bf A}\cup{\bf B}$ of
the sets {\bf A} and {\bf B} can be obtained from the GF $E(z)$ of their
intersection
$$
U(z)=A(z)+B(z)-E(z).
$$

%D
\section{The Axioms of Quantum Mechanics}\label{Axioms}

\newcommand\frAxiomA{Les \'etats du syst\`eme de la m\'ecanique quantique se d\'ecrivent par de vecteurs $|\psi\rangle$ de l'espace hilbertien abstrait.}
\newcommand\frAxiomB{Aux variables dynamiques de la m\'ecanique quantique on fait correspondre des op\'erateurs lin\'eaires et hermitiens $F$ agissant dans l'espace hilbertien des vecteurs d'\'etats.}
\newcommand\frAxiomC{Les seuls r\'esultats possibles des mesures de la variable dynamique donn\'e dans l'\'etat d\'efini du syt\`eme sont les valeurs propres de l'op\'erateur associ\'e $F$.}
\newcommand\frAxiomD{La probabilit\'e $W_\psi(f)$ d'obtenir par mesure la variable dynamique $F$ dans l'\'etat $|\psi\rangle$ la valeur est donn\'e par la formule
	$%\begin{equation}\label{40.5}
	W_\psi(f)=|\langle f\rangle|^2
	$ %\end{equation}
	o\`u $f$ est le vecteur propre de l'op\'erateur $F$ appartenant \`a la valeur propre $f$.}
\newcommand\frAxiomE{Aux coordonn\'ees $x_j$ et aux impulsions $p_j$ ($j$ \'etant le num\'ero du degr\'e de libert\'e) correspondent dans le syst\`eme de la m\'ecanique quantique les op\'erateurs ${\cal X}_j$ et ${\cal P}_j$ satisfaisant aux relations de commutation
	\begin{equation}\label{40.14}
	{\cal X}_j{\cal P}_k-{\cal P}_k{\cal X}_j=\imath \hbar \delta_{jk}{\bf I}
	\end{equation}
	o\`u $\hbar$ est la constante de Planck.}
	
\newtheorem{axiom}{Axiom}
\newcommand\Axiom[2]{\begin{quote}\begin{axiom}\label{#1}{#2}\end{axiom}\end{quote}}

This Appendix reproduces the axioms of quantum theory, as formulated by Chpolski~\CITE{Chpolski}:

\Axiom{frAxiomA}{\frAxiomA}
\Axiom{frAxiomB}{\frAxiomB}
\Axiom{frAxiomC}{\frAxiomC}
\Axiom{frAxiomD}{\frAxiomD}
\Axiom{frAxiomE}{\frAxiomE}

\end{document}